%% file: main.tex
\documentclass{article}

\usepackage[english]{babel}
\usepackage[a4paper,top=2cm,bottom=2cm,left=3cm,right=3cm,marginparwidth=1.75cm]{geometry}

\usepackage{natbib}
\usepackage{amsmath}
\usepackage{amssymb}
\usepackage{amsthm}
\usepackage{mathtools}\mathtoolsset{showonlyrefs=true}
\usepackage{graphicx}
\usepackage{bm}
\usepackage{xcolor}
\usepackage[colorlinks=true, allcolors=blue]{hyperref}
\usepackage{longtable}
\usepackage{booktabs}
\usepackage{multirow}
\usepackage{float}
\usepackage{caption}
\usepackage{subcaption}
\usepackage{tabularx} 

\usepackage{dsfont}
\usepackage[ruled,vlined]{algorithm2e}

\usepackage{color}
\usepackage[normalem]{ulem}

\newcounter{rem}
\newtheorem{remark}[rem]{Remark}

\newtheorem{assumption}{Assumption}

\providecommand{\keywords}[1]{\textbf{\textit{Keywords---}} #1}

\makeatletter
\newsavebox{\hangingbox}
\newcommand{\newhanging}{\par\parindent=\wd\hangingbox}
\newcommand{\SetKwHanging}[2]{%
  \algocf@newcommand{#1}[1]{%
    \sbox\hangingbox{\hbox{\KwSty{#2}\algocf@typo\ }}%
    {\let\\\newhanging\hangindent=\wd\hangingbox\hangafter=1\unhbox\hangingbox##1}%
  }%
}%
\makeatother
\SetKwHanging{KwInput}{Input:}
\SetKwHanging{KwInitialization}{Initialization:}
\SetKwHanging{KwFPCA}{Computation of the MFPCA components:}
\SetKwHanging{KwUFPCA}{Computation of the univariate components:}
\SetKwHanging{KwScores}{Computation of scores:}
\SetKwHanging{KwUniScores}{Computation of the univariate scores:}
\SetKwHanging{KwCov}{Computation of the derivatives of the covariances:}
\SetKwHanging{KwEigenDerivatives}{Computation of the derivatives of the eigencomponents:}
\SetKwHanging{KwDerivatives}{Estimate the derivatives:}
\SetKwHanging{KwOutput}{Output:}

\newcommand{\EE}{\mathbb{E}} 
\newcommand{\RR}{\mathbb{R}} 
\newcommand{\NN}{\mathbb{N}} 
\newcommand{\dd}{{\rm d}}

\newcommand{\TT}[1]{\mathcal{T}_{#1}} 
\newcommand{\sLp}[1]{\mathcal{L}^{2}\left(#1\right)} 
\newcommand{\HH}{\mathcal{H}} 

\newcommand{\pointt}{\mathbf{t}} 
\newcommand{\points}{\mathbf{s}} 

\newcommand{\Xtrunc}[1]{X_{\lceil #1 \rceil}} 
\newcommand{\Xp}[1]{X^{[#1]}} 

\newcommand{\Cpq}[3]{C_{#3}^{[#1, #2]}}  
\newcommand{\Gammap}[3]{(\Gamma_{#3} #1)^{[#2]}}  

\newcommand{\fp}[1]{f^{[#1]}}

\newcommand{\ptl}[1]{\partial^{#1}} 
\newcommand{\derivt}{\frac{\ptl{d}}{\partial \pointt^d}} 
\newcommand{\derivtp}[1]{\frac{\ptl{d}}{\partial t_{#1}^d}} 
\newcommand{\derivsp}[1]{\frac{\ptl{d}}{\partial s_{#1}^d}} 

\newcommand{\psip}[1]{\psi^{[#1]}} 
\newcommand{\hpsi}{\widehat{\psi}_k}

\newcommand{\hrhoi}{\widehat{\rho}_{ik}}
\newcommand{\hnu}{\widehat{\nu}_k}

\newcommand{\nudk}[2]{\nu_{#1, #2}}
\newcommand{\tnudk}[2]{\widetilde{\nu}_{#1, #2}}
\newcommand{\hnudk}[2]{\widehat{\nu}_{#1, #2}}

\newcommand{\rhodk}[2]{\rho_{#1, #2}}
\newcommand{\trhodk}[2]{\widetilde{\rho}_{#1, #2}}
\newcommand{\hrhodk}[2]{\widehat{\rho}_{#1, #2}}

\newcommand{\psidk}[2]{\psi_{#1, #2}}
\newcommand{\tpsidk}[2]{\widetilde{\psi}_{#1, #2}}
\newcommand{\hpsidk}[2]{\widehat{\psi}_{#1, #2}}

\newcommand{\psidkp}[3]{\psi^{[#1]}_{#2, #3}} 
\newcommand{\tpsidkp}[3]{\widetilde{\psi}^{[#1]}_{#2, #3}} 

\newcommand{\Mi}[1]{\mathrm{M}_{#1}}
\newcommand{\Mip}[1]{\mathrm{M}_i^{[#1]}}
\newcommand{\Tim}{\mathrm{T}_{i, \mathrm{m}}}
\newcommand{\Timp}[1]{\mathrm{T}_{i, m_{#1}}^{[#1]}}
\newcommand{\Xip}[1]{X_{#1}^{[p]}} 
\newcommand{\Yim}{Y_{i, \mathrm{m}}} 
\newcommand{\Yimp}[1]{Y_{i, m_{#1}}^{[#1]}} 
\newcommand{\eim}{\varepsilon_{i, \mathrm{m}}} 

\newcommand{\hderivXi}[1]{\widehat{\ptl{d}X}_{#1}} 
\newcommand{\hCd}[1]{\widehat{C}_d} 
\newcommand{\hCpq}[3]{\widehat{C}_{#3}^{[#1, #2]}}  

\DeclareMathOperator{\Var}{Var}

\newcommand\restr[2]{{ %
  \left.\kern-\nulldelimiterspace  %
  #1  %
  \vphantom{\big|}  %
  \right|_{#2}  %
}}

\title{Derivative Estimation of Multivariate Functional Data}
\author{%
Yueyun Zhu\thanks{School of Mathematical and Statistical Sciences, University of Galway, Ireland \href{mailto:yueyun.zhu@universityofgalway.ie}{yueyun.zhu@universityofgalway.ie}}
\and
Steven Golovkine\thanks{MACSI, Department of Mathematics and Statistics, University of Limerick, Ireland \href{mailto:steven.golovkine@ul.ie}{steven.golovkine@ul.ie}}
\and
Norma Bargary\thanks{MACSI, Department of Mathematics and Statistics, University of Limerick, Ireland \href{mailto:norma.bargary@ul.ie}{norma.bargary@ul.ie}}
\and
Andrew J. Simpkin\thanks{School of Mathematical and Statistical Sciences, University of Galway, Ireland \href{mailto:andrew.simpkin@nuigalway.ie}{andrew.simpkin@nuigalway.ie}}
}
\date{\today}

\begin{document}
\maketitle

\begin{abstract}
Existing approaches for derivative estimation are restricted to univariate functional data. We propose two methods to estimate the principal components and scores for the derivatives of multivariate functional data. As a result, the derivatives can be reconstructed by a multivariate Karhunen-Loève expansion. The first approach is an extended version of multivariate functional principal component analysis (MFPCA) which incorporates the derivatives, referred to as derivative MFPCA (DMFPCA). The second approach is based on the derivation of multivariate Karhunen-Loève (DMKL) expansion. We compare the performance of the two proposed methods with a direct approach in simulations. The simulation results indicate that DMFPCA outperforms DMKL and the direct approach, particularly for densely observed data. We apply DMFPCA and DMKL methods to coronary angiogram data to recover derivatives of diameter and quantitative flow ratio. We obtain the multivariate functional principal components and scores of the derivatives, which can be used to classify patterns of coronary artery disease.
\end{abstract}


\input{main/introduction}

\input{main/methodology}

\input{main/inference}

\input{main/estimation}

\input{main/simulation}
 
\input{main/application}

\input{main/discussion}


\appendix

\section*{Acknowledgment}

This work was supported by Science Foundation Ireland (SFI) grant 19/FFP/7002.

\bibliographystyle{apalike}
\bibliography{./biblio.bib}

\end{document}

%% file: main/introduction.tex
\section{Introduction} 
\label{sec:introduction}

With the development of wearable monitoring devices, sensors, and neuroimaging technologies, increasingly large and complex datasets are being recorded. Functional data analysis (FDA) has become increasingly important in this context. Early studies of FDA have been explored by \cite{rao1958some,ramsay1982data} and \cite{ramsay1991some}. More recent works include, e.g., \cite{ramsay2002applied,Ramsay2005FDA,manteiga2007statistics} and \cite{kokoszka2017introduction}. 

Many functional data do not change linearly over time. When analysing such nonlinear trajectories, the primary interest is often on the rate of change, which can be directly assessed through estimating the velocity and acceleration (i.e. derivative estimation). In FDA, derivative estimations have been applied to many areas, such as online auction prices \citep{wang2008explaining, liu2009estimating}, option prices  \citep{grithFunctionalPrincipalComponent2018}, prepubertal growth speed \citep{hall2009estimation} and blood pressure during pregnancy \citep{simpkin2018derivative}. 

Numerical differentiation \citep{flannery1992numerical} is a traditional way to calculate raw derivatives as the sequential change in response. However, unless the response is collected at high frequency with a smooth trajectory, such simple differences are not reliable. On the other hand, kernel-based methods are commonly used to fit curves and estimate their derivative. Early works discussing kernel-based derivative estimation include \cite{gasser1984estimating} and \cite{muller1987bandwidth}. \cite{fan1995local} and \cite{fan1996local} investigated the generalisation of local polynomial fitting with kernel weights, which makes it straightforward to estimate derivatives. Moreover, B-splines \citep{de1972calculating}, smoothing splines \citep{ramsay1982data,zhou2000derivative}, P-splines \citep{cao2012estimating,simpkin2013additive,eilers2021practical,hernandez2023derivative} and mixed effect models with splines \citep{simpkin2018derivative} have been utilised for derivative estimation.

Functional principal component analysis (FPCA) \citep{silverman1996smoothed, james2000principal, yao2005functional} is a key dimension reduction tool for functional data. FPCA aims to represent the infinite-dimensional functional data into the Karhunen-Loève expansion with a set of orthogonal functional principal components (FPCs) and FPC-scores. FPCs characterise the modes of variation of the functional data and the FPC-scores are uncorrelated random variables serving as coefficients of FPCs. Following this, \cite{liu2009estimating} proposed a method to estimate the derivatives of FPCs, which are then plugged into the derivation of Karhunen-Loève expansion to obtain estimates of the derivatives of functional data. However, because the derivatives of FPCs are not orthogonal, it leads to a suboptimal expansion. \cite{dai2018derivative} aimed to estimate orthogonal FPCs for the derivatives of functional data, and employed a best linear unbiased prediction (BLUP) to predict the corresponding scores. In this case, the derivatives of functional data are expressed in their Karhunen-Loève representation (the optimal expansion). \cite{grithFunctionalPrincipalComponent2018} extended the implementation of FPCA to derivatives in a multi-dimensional domain. They proposed two approaches to recover the derivatives and estimate their eigencomponents. One is based on obtaining the Karhunen-Loève expansion of the derivatives, and the other starts from differentiating the Karhunen-Loève expansion of the original data. 

The aforementioned methods for derivative estimation are restricted to univariate cases. In many applications, data are collected in a multivariate form with multiple features per observation unit. Multivariate functional principal component analysis (MFPCA) \citep{chiou2014multivariate,happ2018multivariate} allows one to capture the joint variation in different features. We are interested in extending MFPCA to derivatives, including the recovery of derivatives and the estimation of their eigencomponents and scores. A simple example is children's gait cycles \citep{Ramsay2005FDA}. The first derivatives of knee and hip angles represent the angular velocity. The eigenfunctions of these derivatives capture the joint variation in the rates of angle changes, and the scores summarise such variation for each gait.

In this paper, we aim to extend MFPCA to estimate the principal components and scores for the derivatives of multivariate functional data. As a result, the derivatives can then be reconstructed by a multivariate Karhunen-Loève expansion. We present two approaches. The first approach considers each univariate feature separately in a preliminary step. It starts with the estimation of eigencomponents of the derivatives for each univariate feature.  Following \cite{happ2018multivariate}, the eigencomponents of the derivatives for multivariate functional data can be linked to their univariate counterparts. The second approach considers all features from the beginning, by expressing them in a multivariate Karhunen-Loève expansion. Later steps for estimating the eigencomponents of the derivatives for multivariate functional data are based on the derivation of this multivariate Karhunen-Loève expansion. 

The paper is organised as follows. In Section~\ref{sec:methodology}, we introduce two approaches to estimate the eigencomponents and scores of the derivatives for multivariate functional data. Section~\ref{sec:inference} provides the inference and Section~\ref{sec:estimation_procedures} presents the estimation procedures for the two approaches. A simulation study to evaluate the two approaches is provided in Section~\ref{sec:simulation}. Section~\ref{sec:application} shows an application to coronary angiogram data, with the potential to guide cardiologists in identifying the optimal location for stent insertion and classifying coronary artery disease patterns. The paper concludes with a discussion in Section~\ref{sec:discussion}.

%% file: main/methodology.tex
\section{Methodology} 
\label{sec:methodology}

We consider independent realisations of $X = (\Xp{1}, \dots, \Xp{P})^\top$, $P \geq 1$, a centered vector-valued stochastic process which consists of $P$ trajectories. (Here and in the following, for any matrix $A$, $A^\top$ denotes its transpose.) For each $1 \leq p \leq P$, the feature $\Xp{p}: \TT{p} \rightarrow \RR$ is assumed to belong to  $\sLp{\TT{p}}$, with $\TT{p} \subset \RR$. Let $\TT{} \coloneqq \TT{1} \times \dots \times \TT{P}$ be the domain of $X$ and denote by $\pointt \coloneq (t_1, \ldots, t_P)$ an element of $\TT{}$. The process $X(\pointt) = (\Xp{1}(t_1), \dots, \Xp{P}(t_P))^\top$ is then defined on $\HH \coloneqq \sLp{\TT{1}} \times \dots \times \sLp{\TT{P}}$. We assume that each feature of the process has finite second moment, i.e., $\int_{\TT{p}} \EE(\{\Xp{p}(t_p)\}^2) \dd t_p < \infty, p = 1, \dots, P$. Let $C$ denotes the $P \times P$ matrix-valued covariance function of $X$ which is defined as
\begin{equation}\label{eq:covariance_function}
    C(\points, \pointt) \coloneqq \EE\left(\left\{X(\points)\right\} \left\{X(\pointt)\right\}^{\top}\right), \quad \points, \pointt \in \TT{}.
\end{equation}
More precisely, for $1 \leq p, q \leq P$, the $(p, q)$th entry of the matrix $C(\points, \pointt)$ is the covariance function between the $p$th and the $q$th features of the process $X$:
\begin{equation}\label{eq:covariance_function_components}
	\Cpq{p}{q}{}(s_p, t_q) \coloneqq \EE\left(\left\{\Xp{p}(s_p)\right\}\left\{\Xp{q}(t_q)\right\}\right), \quad s_p \in \TT{p}, t_q \in \TT{q}.
	\end{equation}
The covariance operator $\Gamma : \HH \rightarrow \HH$ of $X$ is defined as an integral operator with kernel $C$. For $f \in \HH$, the $p$th component of $\Gamma f(\cdot)$ is given by
\begin{equation}\label{eq:covariance_operator_components}
	\Gammap{f}{p}{}(t_p) \coloneqq \sum_{q = 1}^{P} \int_{\TT{q}} \Cpq{p}{q}{}(t_p,s_q)\fp{q}(s_q) \dd s_q, \quad t_p \in \TT{p}.
\end{equation}

Assuming that the covariance operator $\Gamma$ is a compact positive operator on $\HH$, the theory of the Hilbert-Schmidt operator (see, e.g., \cite{reedMethodsModernMathematical1980}) states the existence of an orthonormal basis $\{\psi_{k}\}_{k \geq 1}$ and a set of real numbers $\{\nu_{k}\}_{k \geq 1}$ such that $\nu_{1} \geq \nu_{2} \geq \dots \geq 0$ satisfying
\begin{equation}\label{eq:eigenequations}
    \Gamma \psi_{k}(\cdot) = \nu_{k}\psi_{k}(\cdot), \quad\text{and}\quad \nu_{k} \longrightarrow 0 \quad\text{as}\quad k \longrightarrow \infty.
\end{equation}
The set $\{\nu_{k}\}_{k \geq 1}$ contains the eigenvalues of the covariance operator $\Gamma$ and $\{\psi_{k}\}_{k \geq 1}$ contains the associated eigenfunctions. Note that the $\psi_{k}$ are elements of $\HH$, and thus are vector-valued functions. The process $X$ allows for the multivariate Karhunen-Loève decomposition \citep{happ2018multivariate}
\begin{equation}\label{eq:kl_multi}
    X(\pointt) = \sum_{k = 1}^\infty \rho_{k} \psi_{k}(\pointt), \quad \pointt \in \TT{}, \quad\text{where}\quad \rho_k = \sum_{p = 1}^P \int_{\TT{p}} \Xp{p}(t_p)\psip{p}_k(t_p) \dd t_p
\end{equation}
are the projections of the curve onto the eigenfunctions, with $\EE(\rho_{k}) = 0$, $\Var(\rho_{k}) = \nu_{k}$ and $\EE(\rho_{k}\rho_{l}) = 0$ for $k \neq l$. The eigenfunctions $\psi_{k}$ are the multivariate functional principal components (MFPCs) and the coefficients $\rho_{k}$ are the multivariate functional principal component scores (MFPC-scores).

Let $f \in \HH$. The $d$th derivative of $f$ is given by
\begin{equation}\label{eq:derivatives_f}
    \derivt f(\pointt) \coloneqq \left(\derivtp{1} \fp{1}(t_1), \dots, \derivtp{P} \fp{P}(t_P)\right)^\top, \quad \pointt \in \TT{}.
\end{equation}
When there is no ambiguity, we use $\ptl{d} f(\pointt)$ and $\ptl{d} \fp{p}(t_p)$ to represent the $d$th derivative of the function $f$ with respect to $\pointt$ and of the $p$th feature of $f$ with respect to $t_p$ such that
\begin{equation}
\ptl{d} f(\pointt) \coloneqq \derivt f(\pointt) \quad\text{and}\quad \ptl{d} \fp{p}(t_p) \coloneqq \derivtp{p} \fp{p}(t_p).
\end{equation}
\begin{remark}
By definition, $f(\pointt) = \ptl{0} f(\pointt)$ and $\fp{p}(t_p) = \ptl{0} \fp{p}(t_p)$. 
\end{remark}
\begin{assumption}\label{assumption}
Throughout the paper, we assume that, 
\begin{enumerate}
	\item for $d \in \NN$, the $d$th derivative of the process $X$ exists almost surely and that $\ptl{d} X(\pointt) \in \HH$,
	\item for all $p = 1, \dots, P$, $\ptl{d}\ptl{d}\Cpq{p}{p}{}(s_p, t_p)$ exists and is continuous, where
	\begin{equation}
	\ptl{d} \ptl{d} \Cpq{p}{p}{}(s_p, t_p) \coloneq \derivtp{p}\derivsp{p}\EE\left(\left\{\Xp{p}(s_p)\right\}\left\{\Xp{p}(t_p)\right\}\right).
	\end{equation}
\end{enumerate}
\end{assumption}

We propose two approaches for modeling the derivatives $\ptl{d} X$ when $d > 0$. On the one hand, considering Assumption~\ref{assumption}, by a direct extension of \cite[Th.~2]{kadota1967differentiation} to vector-valued stochastic processes, we have that
$\ptl{d} \psi_{k}(\pointt)$ exists and is continuous for each $k$, and the term-by-term differentiability of the Karhunen-Loève expansion \eqref{eq:kl_multi},
\begin{equation}\label{eq:DermultiKL}
	\ptl{d} X(\pointt) = \sum_{k = 1}^{\infty} \rho_{k} \ptl{d}\psi_{k}(\pointt), \quad t \in \TT{},
\end{equation}
where $\rho_k$ are the MFPC-scores defined in \eqref{eq:kl_multi} and $\ptl{d} \psi_k(\pointt)$ are the $d$th derivative of MFPCs $\psi_k(\pointt)$ with respect to $\pointt$. The series converges uniformly in $\pointt$ to $\ptl{d} X(\pointt)$. The derivation of the multivariate Karhunen-Loève expansion (DMKL) allows us to estimate the derivatives of the process $X$. For each $k$, the derivative of the eigenfunction $\psi_k(\pointt)$, $\ptl{d}\psi_k(\pointt)$, is the solution to $\ptl{d}(\Gamma \psi_k)(\pointt) = \nu_k \ptl{d} \psi_k(\pointt)$.

\begin{remark}\label{rem:dmkl}
Here, the derivatives $\ptl{d}\psi_k(\cdot)$ are not the orthogonal eigenfunctions of the derivatives of the process $X$, $\ptl{d} X(\cdot)$.
\end{remark}

On the other hand, let $C_d$ denote the $P \times P$ matrix-values covariance function of $\ptl{d} X(\pointt)$ which is defined as:
\begin{equation}\label{eq:covariance_function_derivatives}
	C_d(\points, \pointt) \coloneqq \EE\left(\left\{\ptl{d} X(\points)\right\}\left\{\ptl{d} X(\pointt)\right\}^{\top}\right), \quad \pointt, \points \in \TT{}.
\end{equation}
More precisely, for $1 \leq p, q \leq P$, the $(p, q)$th entry of the matrix $C_d(\pointt, \points)$ is the covariance function between the $p$th and the $q$th features of the process $\ptl{d} X(\pointt)$:
\begin{equation}\label{eq:covd_pq}
	\Cpq{p}{q}{d}(s_p, t_q) \coloneqq \EE\left(\left\{\ptl{d} \Xp{p}(s_p)\right\}\left\{\ptl{d}\Xp{q}(t_q)\right\}\right), \quad s_p \in \TT{p}, t_q \in \TT{q}.
\end{equation}

Consider the $p$th feature, the partial derivative of covariance $\ptl{d} \ptl{d} \Cpq{p}{p}{}$ can be further represented as \cite[]{kadota1967differentiation,dai2018derivative}:
\begin{equation}\label{eq:PartialDerCov}
	\begin{aligned}
		\ptl{d} \ptl{d} \Cpq{p}{p}{}(s_p, t_p)&=
		\derivtp{p}\derivsp{p}\EE\left(\left\{\Xp{p}(s_p)\right\}\left\{\Xp{p}(t_p)\right\}\right)\\
		&=\EE\left(\left\{\ptl{d} \Xp{p}(s_p)\right\}\left\{\ptl{d}\Xp{p}(t_p)\right\}\right)\\
		&= \Cpq{p}{p}{d}(s_p, t_p), \quad s_p, t_p \in \TT{p},
	\end{aligned}
\end{equation}
where the interchange of expectation and differentiation holds under Assumption~\ref{assumption} and additional regularity conditions, such as $\Xp{p}$ and $\ptl{d} \Xp{p}$ having finite second moment. 

The covariance operator $\Gamma_d : \HH \rightarrow \HH$ of $X$ is defined as an integral operator with kernel $C_d$. Assuming that the covariance operator $\Gamma_d$ is a compact positive operator on $\HH$, it exists an orthonormal basis $\{\psidk{d}{k}\}_{k \leq 1}$ and a set of real numbers $\{\nudk{d}{k}\}_{k \geq 1}$ such that $\nudk{d}{1} \geq \nudk{d}{2} \geq \dots \geq 0$ satisfying
\begin{equation}
	\Gamma_d\psidk{d}{k}(\cdot) = \nudk{d}{k}\psidk{d}{k}(\cdot), \quad \nudk{d}{k} \longrightarrow 0 \quad\text{as}\quad k \longrightarrow \infty
\end{equation}
The set $\{\nudk{d}{k}\}_{k \geq 1}$ contains the eigenvalues of the covariance operator $\Gamma_d$ and $\{\psidk{d}{k}\}_{k \geq 1}$ contains the associated eigenfunctions.
\begin{remark}
The eigenfunction $\psidk{d}{k}$ denotes the $k$th eigenfunction of the covariance operator $\Gamma_d$ and not the $d$th derivative of $\psi_{k}$, i.e., $\psidk{d}{k}(\cdot) \neq \ptl{d} \psi_k(\cdot)$. 	
\end{remark}
Similarly to the process $X$, the process $\ptl{d} X$ can be expanded using the multivariate Karhunen-Loève decomposition
\begin{equation}\label{eq:kl_deriv_multi}
	\ptl{d} X(\pointt) = \sum_{k = 1}^{\infty}\rhodk{d}{k}\psidk{d}{k}(\pointt), \quad \pointt \in \TT{}, \quad\text{where}\quad \rhodk{d}{k} = \sum_{p = 1}^P \int_{\TT{p}} \ptl{d}\Xp{p}(t_p)\psidkp{p}{d}{k}(t_p) \dd t_p,
\end{equation}
with $\EE(\rhodk{d}{k}) = 0$, $\Var(\rhodk{d}{k}) = \nudk{d}{k}$ and $\EE(\rhodk{d}{k}\rhodk{d}{l}) = 0$ for $k \neq l$. The eigenfunctions $\psidk{d}{k}$ are the derivative MFPCs (DMFPCs) and the coefficients $\rhodk{d}{k}$ are the derivative MFPC-scores (DMFPC-scores). The derivative multivariate functional principal component analysis (DMFPCA) consists of expanding $\ptl{d} X$ into \eqref{eq:kl_deriv_multi}.

\begin{remark}\label{rem:deriv0}
In particular, when $d = 0$, we have $\Gamma_0 \equiv \Gamma$, $C_0 \equiv C$, $\nudk{0}{k} \equiv \nu_k$, $\psidk{0}{k} \equiv \psi_k$ and $\rhodk{0}{k} \equiv \rho_{k}$. Additionally, \eqref{eq:kl_deriv_multi} reduces to $X(\pointt)=\sum_{k=1}^{\infty}\rho_k\psi_k(\pointt)$ which corresponds to MFPCA.
\end{remark}

In applications, it is not feasible to use an infinite number of terms in~\eqref{eq:kl_multi} and~\eqref{eq:kl_deriv_multi} to represent the process $X$ and its derivatives $\ptl{d} X$, so the representations must be truncated. Let 
\begin{equation}\label{eq:kl_trunc}
	\Xtrunc{K}(\pointt) = \sum_{k = 1}^K \rho_{k} \psi_{k}(\pointt) \quad\text{and}\quad \ptl{d}\Xtrunc{K}(\pointt) = \sum_{k = 1}^{K}\rhodk{d}{k}\psidk{d}{k}(\pointt), \quad \pointt \in \TT{},
\end{equation}
be the truncated Karhunen-Loève expansion to $K$ terms of $X$ and $\ptl{d} X$ respectively. The representation~\eqref{eq:kl_trunc} is the representation which will induce the most accurate truncation for a given truncation number $K$ \cite[Lemma~A.2]{golovkineClusteringMultivariateFunctional2022}. Therefore, the basis defined in~\eqref{eq:kl_deriv_multi} is more efficient to represent the derivatives than the basis defined in~\eqref{eq:DermultiKL} for a given truncation number.


%% file: main/inference.tex
\section{Inference} 
\label{sec:inference}

Let $X_1, \dots, X_N \in \HH$ be independent realisations of the process $X$. For $d > 0$, we assume that the $d$th derivative of each realisation $X_i, i = 1, \dots, N$ exists almost surely and $\ptl{d} X_i \in \HH$. If the independent derivatives $\ptl{d} X_1, \dots, \ptl{d} X_N$ were observed, the ideal estimator of the covariance function would be
\begin{equation}\label{eq:covariance_function_derivatives_tilde}
    \widetilde{C}_d(\points, \pointt) = \frac{1}{N - 1}\sum_{i = 1}^N\left\{\ptl{d} X_i(\points)\right\}\left\{\ptl{d} X_i(\pointt)\right\}^{\top}, \quad \pointt, \points \in \TT{}.
\end{equation}
Using this estimator of the covariance, the ideal approximation of the covariance operator $\Gamma_d$ is $\widetilde{\Gamma}_d$, the integral operator with kernel $\widetilde{C}_d$. The ideal estimators of the eigencomponents of $\ptl{d} X$, $\tnudk{d}{k}$ and $\tpsidk{d}{k}$, are the solutions of the equation $\widetilde{\Gamma}_d\tpsidk{d}{k}(\cdot) = \tnudk{d}{k}\tpsidk{d}{k}(\cdot)$. Finally, we have, for each $i = 1, \dots, N$,
\begin{equation}
    \ptl{d} X_i(\pointt) = \sum_{k = 1}^K \trhodk{d}{ik}\tpsidk{d}{k}(\pointt), \quad t \in \TT{}, \quad\text{where}\quad \trhodk{d}{ik} = \sum_{p = 1}^P \int_{\TT{p}} \ptl{d}\Xp{p}_i(t_p)\tpsidkp{p}{d}{k}(t_p) \dd t_p.
\end{equation}

In applications, the curves are rarely observed without error and never at each value $t \in \TT{}$. We thus consider this more realistic setup. For each observation $i = 1, \dots, N$, let $\Mi{i} = (\Mip{1}, \dots, \Mip{P}) \in \RR^{P}$ be a vector of positive integers representing the number of sampling points of each feature of observation $i$. Let $\Tim = (\Timp{1}, \dots, \Timp{P}), 1 \leq m_p \leq \Mip{p}, 1 \leq p \leq P$, be the random sampling points for the observation $X_i$. These sampling points are obtained as independent realisations of a random variable $\mathbf{T}$ taking values in $\TT{}$. The vectors $\Mi{1}, \dots, \Mi{N}$ represent an independent sample of an integer-valued random vector $\mathbf{M}$ with known expectation. We assume that the realisations of $X$, $\mathbf{M}$ and $\mathbf{T}$ are mutually independent. The observations associated with a realisation $X_i$ consist of the pairs of $(\Yim, \Tim) \in \RR^P \times \TT{}$, where $\mathrm{m} = (m_1, \dots, m_P), 1 \leq m_p \leq \Mip{p}, 1 \leq p \leq P$ and $\Yim$ is defined as
\begin{equation}\label{eq:sampling_model}
    \Yim = X_i(\Tim) + \eim, \quad i = 1, \dots, N,
\end{equation}
where $\eim$ being independent copies of a centered error random vector $\varepsilon \in \RR^{P}$ with finite variance $\sigma^2$. We aim to build feasible version of the covariance $\widetilde{C}_d$, the eigenvalues $\tnudk{d}{k}$ and the eigenfunctions $\tpsidk{d}{k}$ for $d > 0$.


%% file: main/estimation.tex
\section{Estimation procedures} 
\label{sec:estimation_procedures}

We introduce two algorithms for estimating the eigencomponents of $\ptl{d} X$. The first algorithm, based on the idea of DMFPCA, starts from the featurewise estimation of the partial derivatives of the covariance function $\Cpq{p}{p}{}$ and then follows MFPCA to obtain the eigencomponents of $\ptl{d} X$. The second algorithm, based on the idea of DMKL, starts from running MFPCA on the realisations of $X$ and computing the derivatives of MFPCs. We denote by $\hderivXi{}$, a general estimator of an observation $\ptl{d}X$, by $\hCd{d}$, an estimator of the covariance $C_d$ defined in~\eqref{eq:covariance_function_derivatives}, and by $\hnudk{d}{k}$, $\hpsidk{d}{k}$ and $\hrhodk{d}{k}$, the estimators of the eigenvalues $\nudk{d}{k}$, of the eigenfunctions $\psidk{d}{k}$ and the scores $\rhodk{d}{k}$ respectively.

\subsection{Using the derivatives of the covariances} 
\label{sub:using_the_derivatives_of_the_covariances}

This method is based on the estimation of partial derivatives of the univariate covariance functions. We compute, for each $p = 1, \dots, P$, the partial derivatives of the estimated covariance function $\hCpq{p}{p}{}$, $\ptl{d} \ptl{d} \hCpq{p}{p}{}$. Here, $\ptl{d} \ptl{d} \hCpq{p}{p}{}$ is a general estimator of $\ptl{d} \ptl{d} \Cpq{p}{p}{}$ and its choice is left to the reader, such as the finite differences method and local polynomial estimation. Based on \eqref{eq:PartialDerCov}, the spectral decomposition of $\ptl{d} \ptl{d} \hCpq{p}{p}{}$ leads to
\begin{equation}
\ptl{d} \ptl{d} \hCpq{p}{p}{}=\sum_{k=1}^{\infty} \widehat{\lambda}^{[p]}_{d,k} \widehat{\phi}^{[p]}_{d,k}(s_p) \widehat{\phi}^{[p]}_{d,k}(t_p),    
\end{equation}
where $\widehat{\lambda}^{[p]}_{d,k}$ are the eigenvalues and  $\widehat{\phi}^{[p]}_{d,k}$ are the corresponding eigenfunctions. The Karhunen-Loève expansion of $\ptl{d} \Xip{i}$ is then represented as
\begin{equation}
    \widehat{\ptl{d}\Xip{i}}(t_p)= \sum_{k=1}^{K_p}\widehat{\xi}^{[p]}_{d,ik} \widehat{\phi}^{[p]}_{d,k}(t_p),
\end{equation}
where the first $K_p$ eigencomponents are selected. Following \cite{dai2018derivative}, we construct best linear unbiased predictors of the univariate scores $\widehat{\xi}^{[p]}_{d,ik}$ from the observed measurements $\{(\Yim, \Tim)\}_{1 \leq i \leq N}$. Using the relationship between univariate and multivariate decomposition \cite[Prop.~5]{happ2018multivariate}, we can then estimate the multivariate eigencomponents, $\hnudk{d}{k}$, $\hpsidk{d}{k}$, as well as the multivariate scores, $\hrhodk{d}{ik}, i = 1, \dots, N$ without explicitly deriving an estimation of $\ptl{d} X_i$. Algorithm~\ref{alg:derivatives_covariance} describes the procedure. Based on Algorithm~\ref{alg:derivatives_covariance}, the derivatives of $X_i$ are reconstructed by:
\begin{equation}\label{eq:refit_derivatives}
	\widehat{\ptl{d} X_i}( \Tim)=\sum_{k=1}^{K}\hrhodk{d}{ik} \hpsidk{d}{k}( \Tim).
\end{equation}

\begin{algorithm*}
\SetAlgoLined
\KwInput{Sample $\{(\Yim, \Tim)\}_{1 \leq i \leq N}$, order of the derivative $d$, number of principal components to estimate $K$.}

\KwCov{For each $p = 1, \dots, P$, estimate $\ptl{d}\ptl{d}\Cpq{p}{p}{}$ using the sample set $\{(\Yimp{p}, \Timp{p})\}_{1 \leq i \leq N}$.}

\KwUFPCA{For each $p = 1, \dots, P$, estimate the univariate eigencomponents using a spectral decomposition of $\ptl{d}\ptl{d}\Cpq{p}{p}{}$.}

\KwUniScores{For each $p = 1, \dots, P$, estimate the univariate scores following \cite{dai2018derivative}.}

\KwFPCA{Using the relationship between univariate and multivariate decomposition \cite[Prop.~5]{happ2018multivariate}, perform an MFPCA with $K$ components and get the set of eigenvalues $\{\hnudk{d}{k}\}_{1 \leq k \leq K}$ associated with the set of eigenfunctions $\{\hpsidk{d}{k}\}_{1 \leq k \leq K}$ from~\eqref{eq:kl_deriv_multi}.}

\KwScores{For each $i = 1, \dots, N$, estimate the set of scores $\{\hrhodk{d}{ik}\}_{1 \leq k \leq K}$ using~\eqref{eq:kl_deriv_multi}.}

\KwOutput{Estimation of the eigencomponents $\hnudk{d}{k}$, $\hpsidk{d}{k}$ and of the scores $\hrhodk{d}{ik}$.}
\caption{Estimation using DMFPCA.}
\label{alg:derivatives_covariance}
\end{algorithm*}


\subsection{Using the derivatives of the Karhunen-Loève expansion} 
\label{sub:using_the_derivatives_of_the_karhunen_loeve_expansion}

Here, we propose to estimate, for each $i = 1, \dots, N$, $\ptl{d} X_i$ and then estimate their eigencomponents. To do so, we first estimate the eigencomponents of $X$, $\hnu$ and $\hpsi$, and the associated scores $\hrhoi$ for each observation $X_i, i = 1, \dots, N$. We then compute, for each $k$, the derivatives of the estimated eigenfunction $\hpsi$, $\ptl{d}\hpsi$. Here, $\ptl{d}\hpsi$ is a general estimator of $\ptl{d}\psi$ and its choice is left to the reader. However, the eigenfunctions should be smooth and thus a finite differences method may be considered. Using~\eqref{eq:DermultiKL}, we estimate an initial version of the derivatives of the observations. This initial estimation of the derivatives is then used as the input for running MFPCA, which leads to the final eigencomponents and the scores for the derivatives. Algorithm~\ref{alg:dmkl_mfpca} describes the method. Based on Algorithm~\ref{alg:dmkl_mfpca}, the final derivatives of $X_i$ are reconstructed from \eqref{eq:refit_derivatives}.

\begin{algorithm*}
\SetAlgoLined
\KwInput{Sample $\{(\Yim, \Tim)\}_{1 \leq i \leq N}$, order of the derivative $d$, number of principal components to estimate $K$.}

\KwFPCA{Perform an MFPCA with $K$ components using the sample set $\{(\Yim, \Tim)\}_{1 \leq i \leq N}$ and get the set of eigenvalues $\{\hnu\}_{1 \leq k \leq K}$ associated with the set of eigenfunctions $\{\hpsi\}_{1 \leq k \leq K}$ from~\eqref{eq:kl_multi}.}

\KwScores{For each $i = 1, \dots, N$, estimate the set of scores $\{\hrhoi\}_{1 \leq k \leq K}$ using~\eqref{eq:kl_multi}.}

\KwEigenDerivatives{For each $k = 1, \dots, L$, estimate $\ptl{d}\hpsi$.}

\KwDerivatives{For each $i = 1, \dots, N$, estimate $\hderivXi{i}$ using \eqref{eq:DermultiKL}.}

\KwFPCA{Perform an MFPCA with $K$ components on the set $\{\hderivXi{1}, \dots, \hderivXi{N}\}$ and get the set of eigenvalues $\{\hnudk{d}{k}\}_{1 \leq k \leq K}$ associated with the set of eigenfunctions $\{\hpsidk{d}{k}\}_{1 \leq k \leq K}$ from~\eqref{eq:kl_deriv_multi}.}

\KwScores{For each $i = 1, \dots, N$, estimate the set of scores $\{\hrhodk{d}{ik}\}_{1 \leq k \leq K}$ using~\eqref{eq:kl_deriv_multi}.}

\KwOutput{Estimation of the eigencomponents $\hnudk{d}{k}$, $\hpsidk{d}{k}$ and of the scores $\hrhodk{d}{ik}$.}
\caption{Estimation using DMKL.}
\label{alg:dmkl_mfpca}
\end{algorithm*}



%% file: main/simulation.tex
\section{Simulation} 
\label{sec:simulation}

We simulate multivariate functional data from four functions ($P=4$):
\begin{align}
	\Xp{1}(t) &= a + \frac{5}{ct + 10b\exp(-16t)}, \\
	\Xp{2}(t) &= a + c \left(\sin(\pi (b+t)) \right)^3,\\
	\Xp{3}(t) &= a - \cos\left(\frac{ct}{4}(2t - \pi)\right) + 6\exp(-16bt^2), \\
	\Xp{4}(t) &= a + 2c\exp(-14t) + 2\exp(2 (t - b)),
\end{align}
where
\begin{align}
	\begin{pmatrix*}
	a\\b\\c
	\end{pmatrix*} \sim
	\mathcal{N}\left( 
	\begin{pmatrix*}
	0\\0.5\\3.75
	\end{pmatrix*},
	\begin{pmatrix*}	
	1&0.11&0.56\\
	0.11&0.02& 0.08\\
	0.56 &0.08 &0.49
	\end{pmatrix*}
	\right).
\end{align}
We consider the most common case of the first derivative $d=1$ and use the notation $\ptl{} X$ for simplicity. The derivatives of the four functions are:
\begin{align}
	\ptl{} \Xp{1}(t) &= \frac{32b\exp(-16t)-c/5}{(ct/5+2b\exp(-16t))^2}, \\
	\ptl{} \Xp{2}(t) &= 3c\pi \cos\left(\pi(b+t)\right) \left(\sin\left(\pi(b+t)\right)\right)^2,\\
	\ptl{} \Xp{3}(t) &= \left(ct-\frac{c\pi}{4}\right)\sin\left(\frac{ct}{4}(2t-\pi)\right)-192bt\exp(-16bt^2), \\
	\ptl{} \Xp{4}(t) &= -28c\exp(-14t)+4\exp(2(t-b)).
\end{align}

\begin{figure}[H]
	\centering
	\begin{subfigure}{\textwidth}
		\centering
		\includegraphics[scale=0.35]{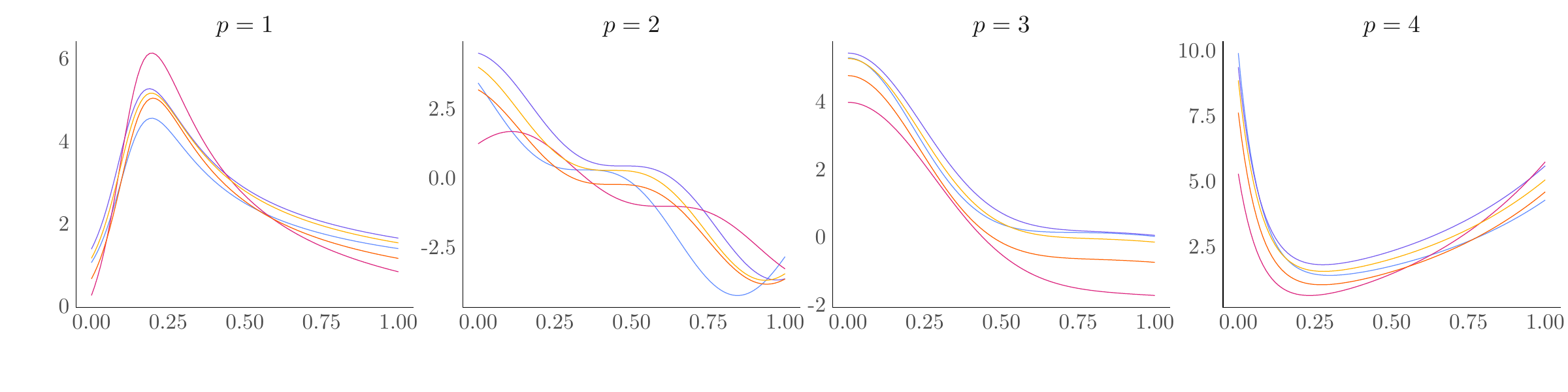}
		\caption{Multivariate functional data.}
	\end{subfigure}
	\\
	\begin{subfigure}{\textwidth}
		\centering
		\includegraphics[scale=0.35]{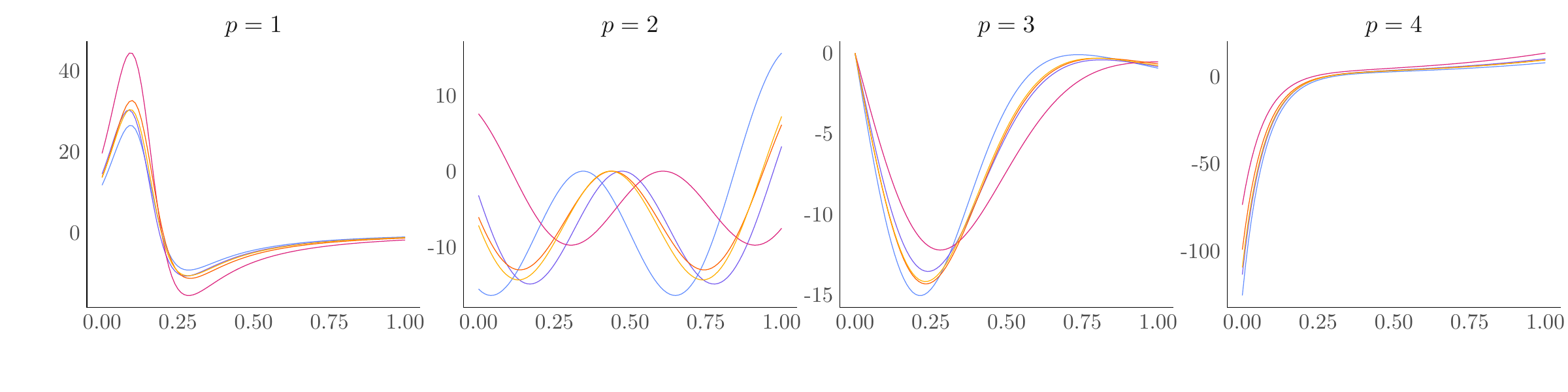}
		\caption{Derivatives of the multivariate functional data.}
	\end{subfigure}
	\caption{(a) Sample of data simulated from the four functions and (b) their corresponding true derivatives.}
	\label{fig:FigTrueData}
\end{figure}

As an illustration, Figure~\ref{fig:FigTrueData} shows a sample of the data simulated from the four functions and their corresponding derivatives. These four features can mimic standard biomedical applications. For instance, the first and the third feature can be used to represent CD4 trajectories in HIV patients undergoing treatment and cognitive function in older age, respectively \citep{simpkin2018derivative}. 

The data are non-centred as shown in Figure~\ref{fig:FigTrueData}. In the case of non-centred data, the choice of mean estimation is left to the reader. Here, we use P-splines \cite{eilers2021practical} to estimate the mean and centre the data before running DMFPCA and DMKL methods.

In this simulation, we have the following four settings:
\begin{enumerate}
	\item Densely observed data without measurement error
	\item Densely observed data with measurement error
	\item Medium-level sparse data with measurement error
	\item High-level sparse data with measurement error
\end{enumerate}

For each setting, we generate $N=100$ random trajectories for each $p=1,\dots,4$ and run the simulation 500 times. For dense cases, the four features are observed on the same time grid on the interval $[0,1]$ consisting of 101 equidistant observations, namely $\Mip{p}=101$ for $p=1,\dots, 4$ and $\Tim \in [0,1]$ for $i=1,\dots,100$. The measurement error is set as $\sigma=0.5$. For sparse cases, the number of observation points $\Mip{p}$ for each $p=1,\dots,4$ is selected from $[50,60]$ and $[10,20]$ for medium and high sparsity, respectively. 

Given that the true derivatives of $X$ are known, we can apply MFPCA to the true derivatives and select the first three components ($K=3$) which account for 93\% of the total variation. The estimated eigencomponents and the scores are treated as the ground truth to evaluate the performance of DMFPCA and DMKL. To evaluate the accuracy of DMFPCA and DMKL, we use relative errors (RE) for eigenvalues, integrated square errors (ISE) for eigenfunctions and relative mean integrated square errors (RMISE) for derivatives, which are defined by:
\begin{equation}
	\begin{aligned}
		\text{RE}(\nu_{d,k}, \widehat{\nu}_{d,k})&=\vert \nu_{d,k}-\hat{\nu}_{d,k} \vert/\nu_{d, k},\\
		\text{ISE}(\psi_{d,k}, \widehat{\psi}_{d,k})&=\sum_{p=1}^{P}\int \left(\psip{p}_{d,k}(t_p)-\widehat{\psi}^{(p)}_{d,k}(t_p)\right)^2dt_p,\\
		\text{RMISE}(\ptl{d} X, \widehat{\ptl{d} X})&=
		\frac{1}{P}\sum_{p=1}^{P}\frac{\sum_{i=1}^{N}\int (\ptl{d} X_i^{(p)}(t_p)-\widehat{\ptl{d}X}_i^{(p)}(t_p))^2dt_p/N}
		{\sum_{i=1}^{N}\int (\ptl{d} X_i^{(p)}(t_p))^2dt_p/N}.
	\end{aligned}
\end{equation}

In addition to the DMFPCA and DMKL methods, a direct approach is utilised to estimate the eigencomponents of the derivatives of the process $X$ as well. This approach first estimates the derivative $\hderivXi{i}$ of $X_i$ for each $i=1,\dots,100$. Here, $\hderivXi{i}$ is a general estimator of $\ptl{d} X_i$ and its choice is left to the reader. We use P-splines in this paper and other options, such as the finite differences method and local polynomial estimation, are also readily available. After that, the eigencomponents and the scores of $\hderivXi{i}$ are estimated by MFPCA, using \cite{happ2018multivariate} or \cite{golovkineUseGramMatrix2023}. Algorithm~\ref{alg:derivatives_obs} describes the approach. Based on Algorithm~\ref{alg:derivatives_obs}, the final derivatives are reconstructed by a Karhunen-Loève expansion as shown in \eqref{eq:refit_derivatives}.

\begin{algorithm*}
	\SetAlgoLined
	\KwInput{Sample $\{(\Yim, \Tim)\}_{1 \leq i \leq N}$, order of the derivative $d$, number of principal components to estimate $K$.}
	
	\KwDerivatives{For each $i = 1, \dots, N$, estimate $\hderivXi{i}$ using the sample $(\Yim, \Tim)$.}
	
	\KwFPCA{Perform an MFPCA with $K$ components on the set $\{\hderivXi{1}, \dots, \hderivXi{N}\}$ and get the set of eigenvalues $\{\hnudk{d}{k}\}_{1 \leq k \leq K}$ associated with the set of eigenfunctions $\{\hpsidk{d}{k}\}_{1 \leq k \leq K}$ from~\eqref{eq:kl_deriv_multi}.}
	
	\KwScores{For each $i = 1, \dots, N$, estimate the set of scores $\{\hrhodk{d}{ik}\}_{1 \leq k \leq K}$ using~\eqref{eq:kl_deriv_multi}.}
	
	\KwOutput{Estimation of the eigencomponents $\hnudk{d}{k}$, $\hpsidk{d}{k}$ and of the scores $\hrhodk{d}{ik}$.}
	\caption{A direct approach to estimate the eigencomponents of $\hderivXi{i}$.}
	\label{alg:derivatives_obs}
\end{algorithm*}

\begin{figure}[H]
	\centering
	\begin{subfigure}{\textwidth}
		\centering
		\includegraphics[scale=0.37]{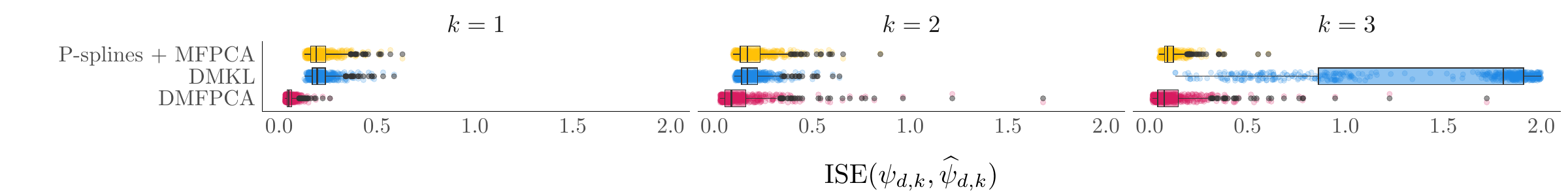}
		\caption{No noise $\sigma = 0$.}
	\end{subfigure}
	\\
	\begin{subfigure}{\textwidth}
		\centering
		\includegraphics[scale=0.37]{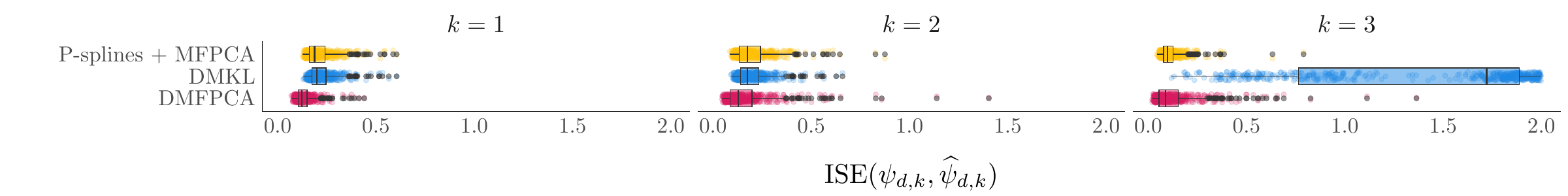}
		\caption{Noise $\sigma = 0.5$}
	\end{subfigure}
	\\
	\begin{subfigure}{\textwidth}
		\centering
		\includegraphics[scale=0.37]{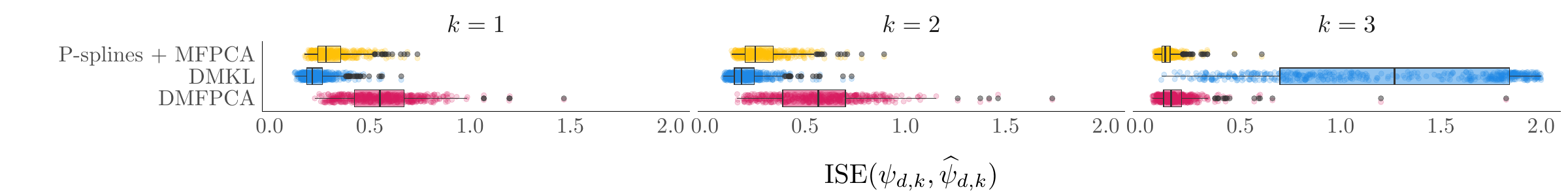}
		\caption{Medium sparsity}
	\end{subfigure}
	\\
	\begin{subfigure}{\textwidth}
		\centering
		\includegraphics[scale=0.37]{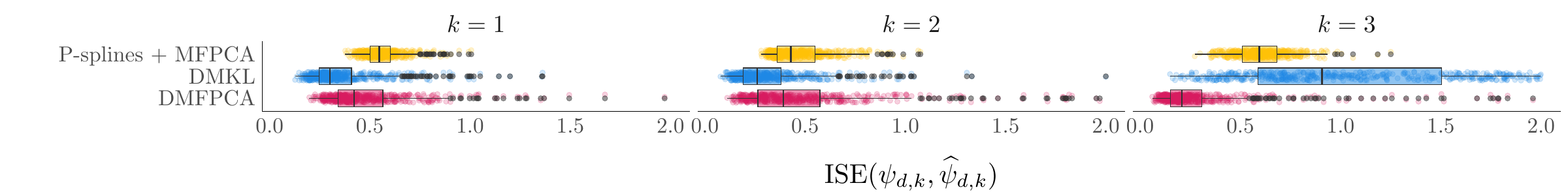}
		\caption{High sparsity.}
	\end{subfigure}
	\caption{ISE for eigenfunctions.}
	\label{fig:ISE_eigenfunctions}
\end{figure}

\begin{figure}[H]
	\centering
	\begin{subfigure}{\textwidth}
		\centering
		\includegraphics[scale=0.37]{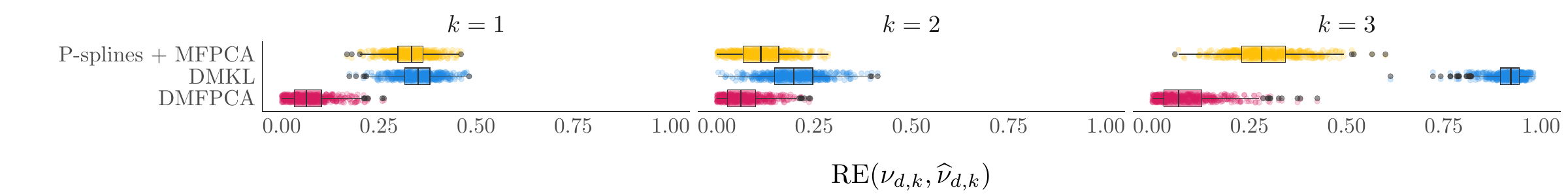}
		\caption{No noise $\sigma = 0$.}
	\end{subfigure}
	\\
	\begin{subfigure}{\textwidth}
		\centering
		\includegraphics[scale=0.37]{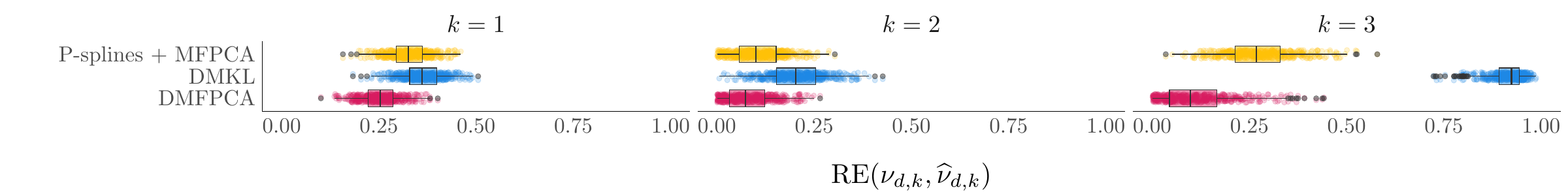}
		\caption{Noise $\sigma = 0.5$}
	\end{subfigure}
	\\
	\begin{subfigure}{\textwidth}
		\centering
		\includegraphics[scale=0.37]{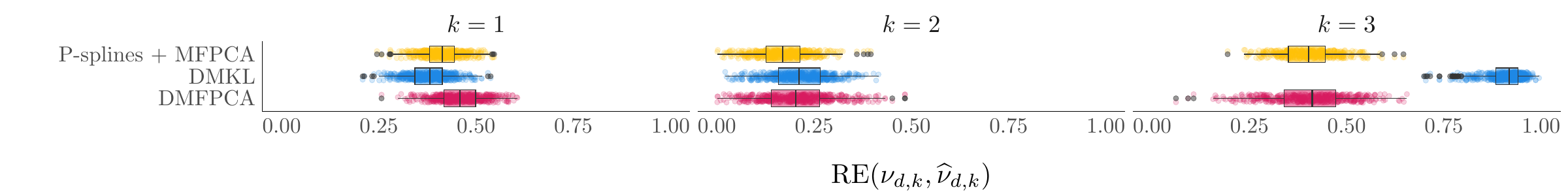}
		\caption{Medium sparsity}
	\end{subfigure}
	\\
	\begin{subfigure}{\textwidth}
		\centering
		\includegraphics[scale=0.37]{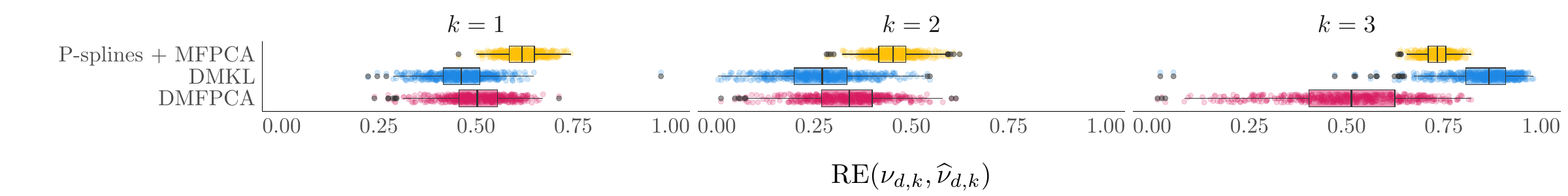}
		\caption{High sparsity.}
	\end{subfigure}
	\caption{RE for eigenvalues.}
	\label{fig:RE_eigenvalues}
\end{figure}

Figure~\ref{fig:ISE_eigenfunctions} and \ref{fig:RE_eigenvalues} present the boxplots of ISE for eigenfunctions and RE for eigenvalues, respectively. The orange bars represent the results obtained from the approach based on Algorithm~\ref{alg:derivatives_obs}, referred to as ``P-splines + MFPCA''. DMKL performs well with the first two eigencomponents, especially for the sparse cases. However, it fails to provide the accurate estimation for the third eigencomponent for all four cases. This is because given the same truncation number, DMKL explains less variance than the true results, as expected from Remark~\ref{rem:dmkl}. When setting $K=3$ in this simulation, the last eigencomponent explains much more variance than estimated by DMKL. Increasing the truncation number could enable DMKL to yield eigencomponents closer to the true result. On the other hand, the figures indicate the estimated eigencomponents from DMFPCA are closer to the ground truth compared to DMKL and P-splines + MFPCA under the dense cases and high sparsity case. For medium sparsity case, DMFPCA yields slightly higher ISE than the P-splines + MFPCA method for all three eigenfunctions. This could be a result of the computation of the derivatives of the covariances, see the second step from Algorithm~\ref{alg:derivatives_covariance}. The estimator of $\ptl{d}\ptl{d}\Cpq{p}{p}{}$ could be improved by using other smoothing methods for the case with medium-level sparsity. In this simulation, we use P-splines with same smoothing parameter to estimate $\ptl{d}\ptl{d}\Cpq{p}{p}{}$ for both medium and high sparsity cases.

\begin{figure}[H]
	\centering
	\begin{subfigure}{0.49\textwidth}
		\centering
		\includegraphics[scale=0.35]{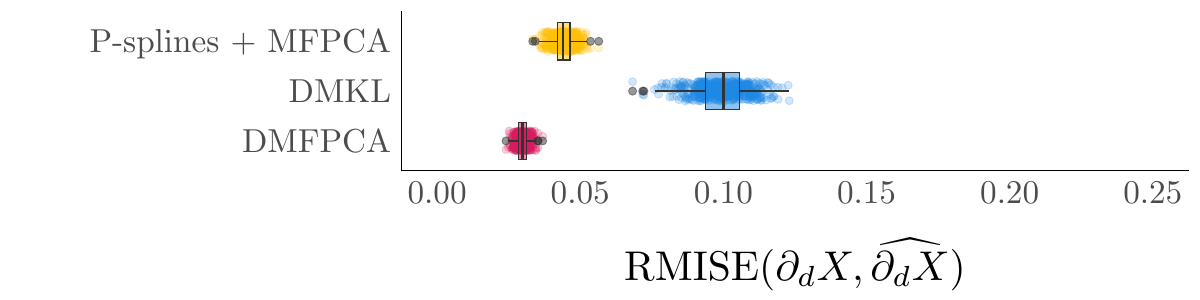}
		\caption{No noise $\sigma = 0$.}
	\end{subfigure}
	\hfill
	\begin{subfigure}{0.49\textwidth}
		\centering
		\includegraphics[scale=0.35]{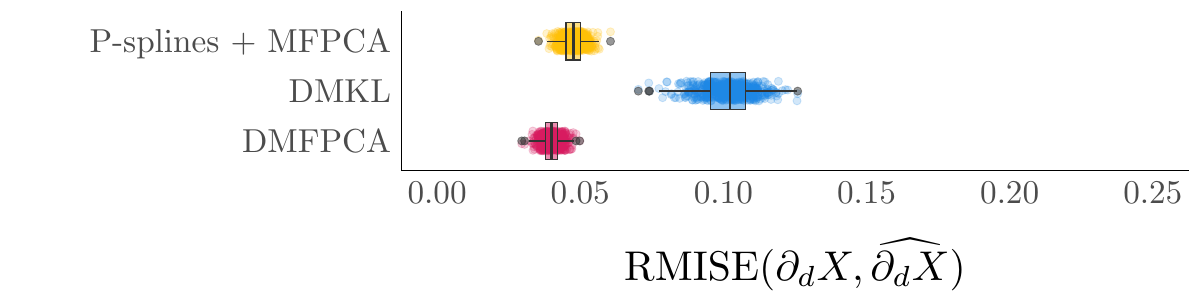}
		\caption{Noise $\sigma = 0.5$}
	\end{subfigure}
	\\
	\begin{subfigure}{0.49\textwidth}
		\centering
		\includegraphics[scale=0.35]{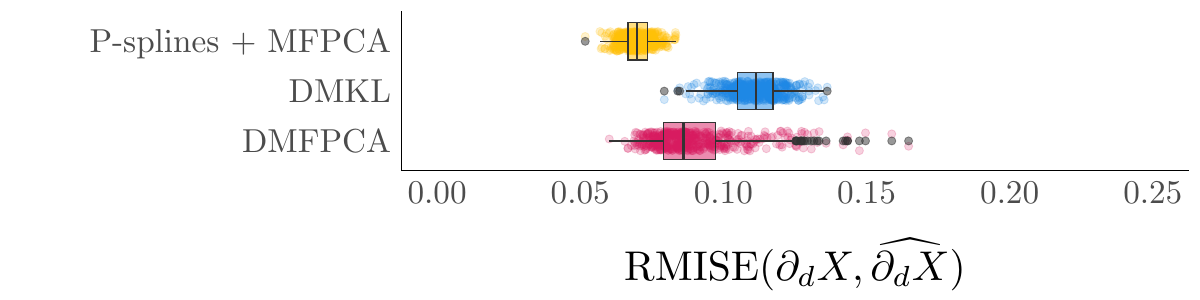}
		\caption{Medium sparsity}
	\end{subfigure}
	\hfill
	\begin{subfigure}{0.49\textwidth}
		\centering
		\includegraphics[scale=0.35]{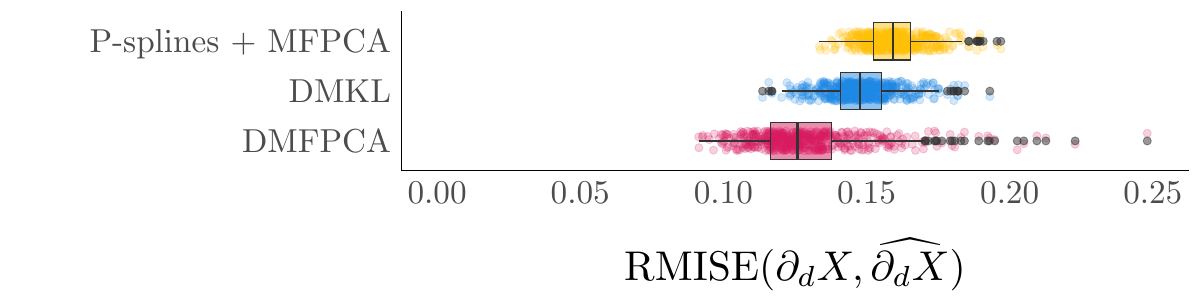}
		\caption{High sparsity.}
	\end{subfigure}
	\caption{RMISE for refitted derivatives.}
	\label{fig:FigRMISE}
\end{figure}

The boxplots of RMISE for refitted derivatives are shown in Figure~\ref{fig:FigRMISE}. It is evident that DMKL yields higher RMISE, particularly for the dense cases. This is because the third eigencomponent estimated by DMKL is not close to the truth, as illustrated above. DMFPCA outperforms the other two methods for dense cases, with average RMISE lower than 5\% for both $\sigma=0$ and $\sigma=0.5$. For a moderate level of sparsity, P-splines + MFPCA provides the lowest RMISE value (around 8\%), whereas DMFPCA and DMKL yields somewhat higher errors (around 10\% and 12\%, respectively). In terms of high sparsity, DMFPCA gives lower RMISE around 12\% compared to DMKL and  P-splines + MFPCA.

Overall, DMFPCA achieves the lowest RMISE, ISE and RE in dense cases, particularly when data are without noise, compared to the other two methods, and remains effective in sparse cases. This indicates that, in practice, DMFPCA should be the preferred choice when data are dense, particularly if data are smooth or pre-smoothed. Unfortunately, with the same truncation number, the performance of DMKL is inferior to that of DMFPCA, with higher ISE and RE for eigencomponents estimation. However, DMKL is still effective to recover the derivatives, with average RMISE lower than 15\% even when data is highly sparse. Furthermore, P-splines + MFPCA works more straightforwardly and can always be used as an alternative backup approach.


%% file: main/application.tex
\section{Application} 
\label{sec:application}

Functional patterns of coronary artery disease (CAD) provide crucial information for planning percutaneous coronary intervention (PCI). The coronary angiograms we use in this section are collected from 139 patients (with 185 vessels) diagnosed with native coronary atherosclerosis that underwent a clinically-driven PCI with at least one bioresorbable vascular scaffold (BRS Absorb; Abbott Vascular, Temecula, USA) implantation between December 2012 and September 2017 at the Verona University Hospital \cite[]{zhu2024validation}. From angiograms, quantitative flow ratio (QFR) and vessels' diameters are obtained, which are essential for assessing the CAD patterns. A focal pattern is defined by a single QFR drop larger than 0.05 over 10mm segment; serial lesions are defined by the presence of two or more focal drops separated by 3 times the reference diameter; diffuse disease is defined by a progressive decline of QFR without clear evidence of a focal drop; and mixed patterns are the combination of previous disease \cite[]{biscaglia2023qfr}.

\begin{figure}[H]
	\centering
	\begin{subfigure}{0.49\textwidth}
		\includegraphics[width=\linewidth]{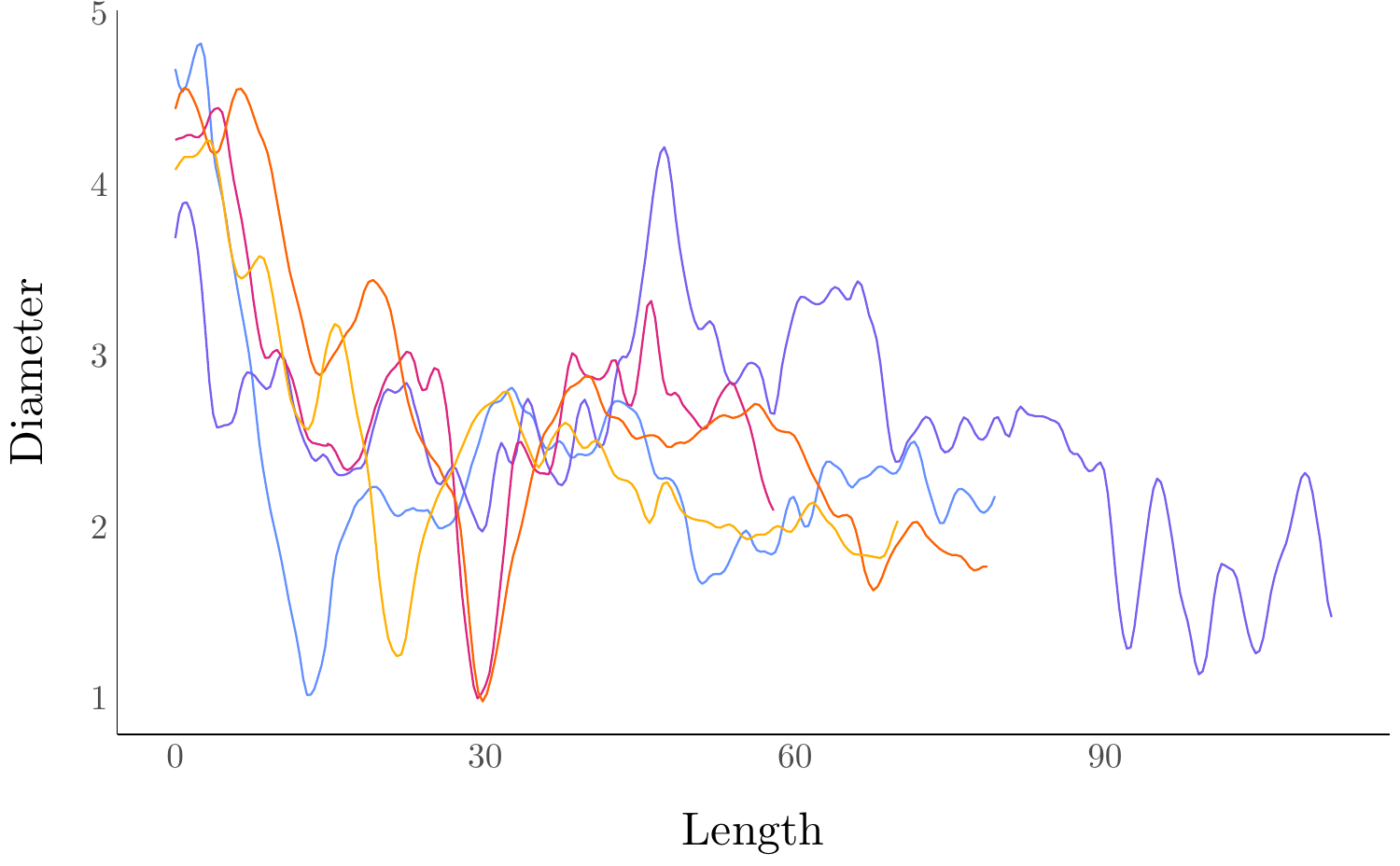}
	\end{subfigure}
	\hfill
	\begin{subfigure}{0.49\textwidth}
		\includegraphics[width=\linewidth]{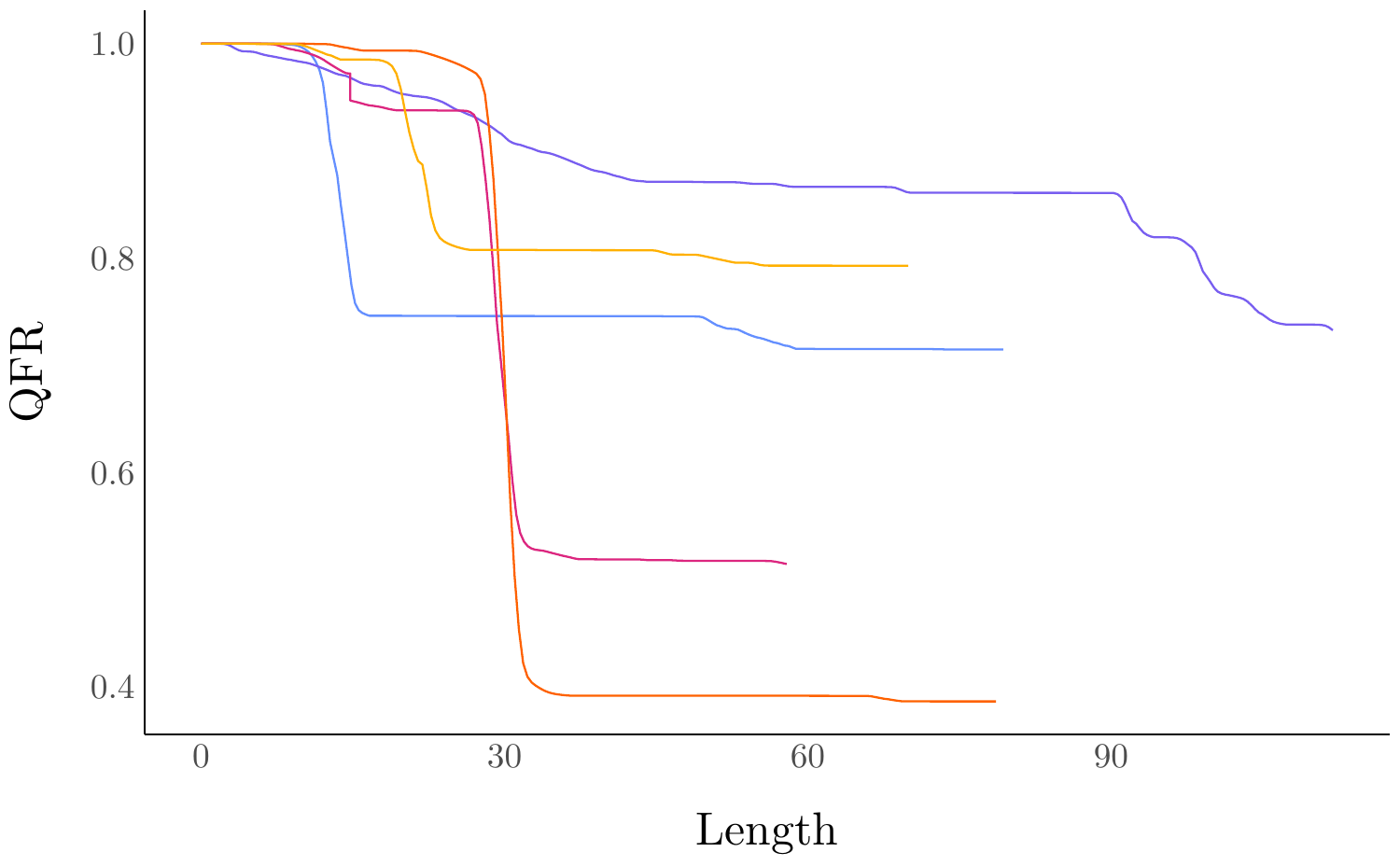}
	\end{subfigure}%
	\\
	\begin{subfigure}{0.49\textwidth}
		\includegraphics[width=\linewidth]{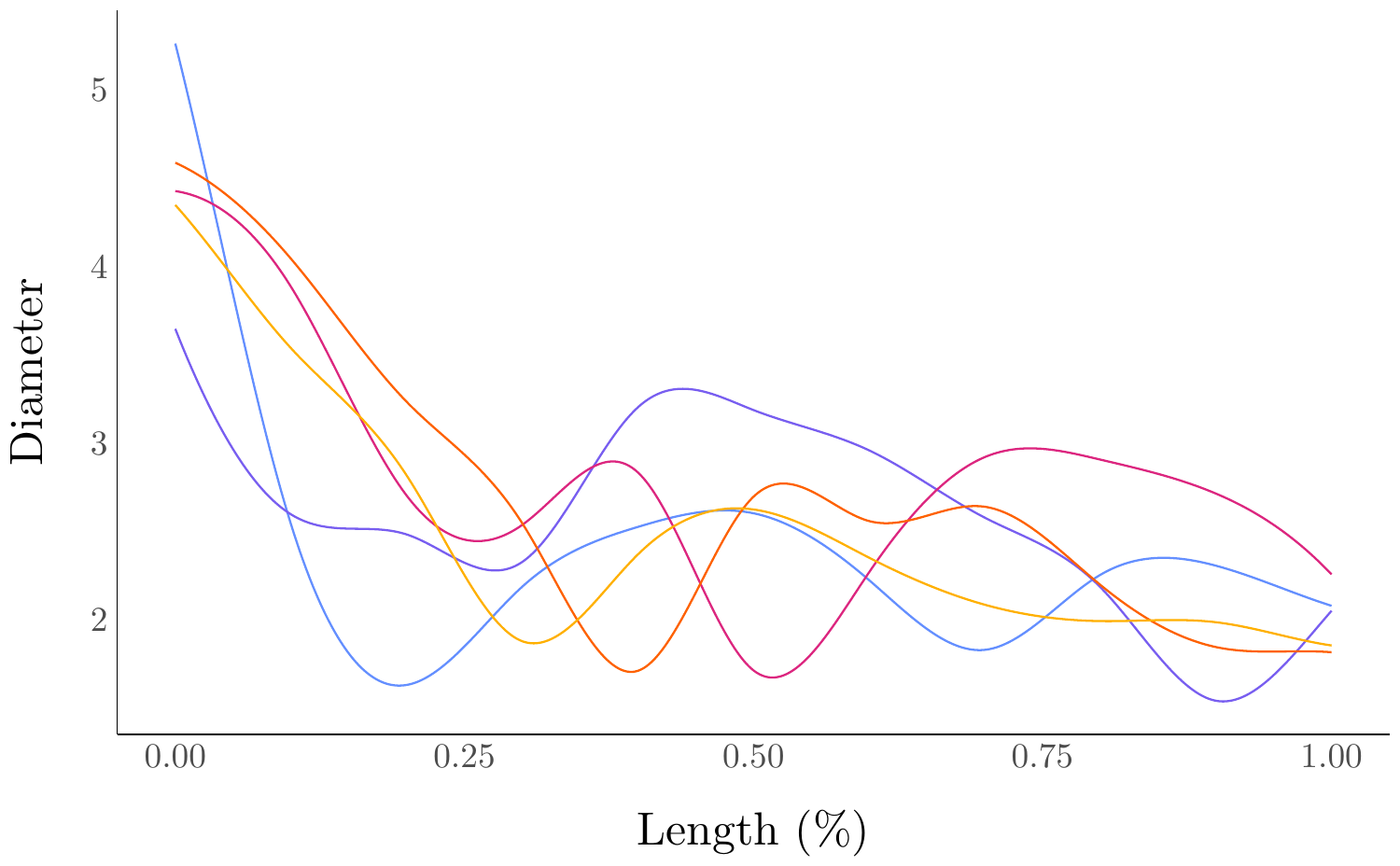}
	\end{subfigure}
	\hfill
	\begin{subfigure}{0.49\textwidth}
		\includegraphics[width=\linewidth]{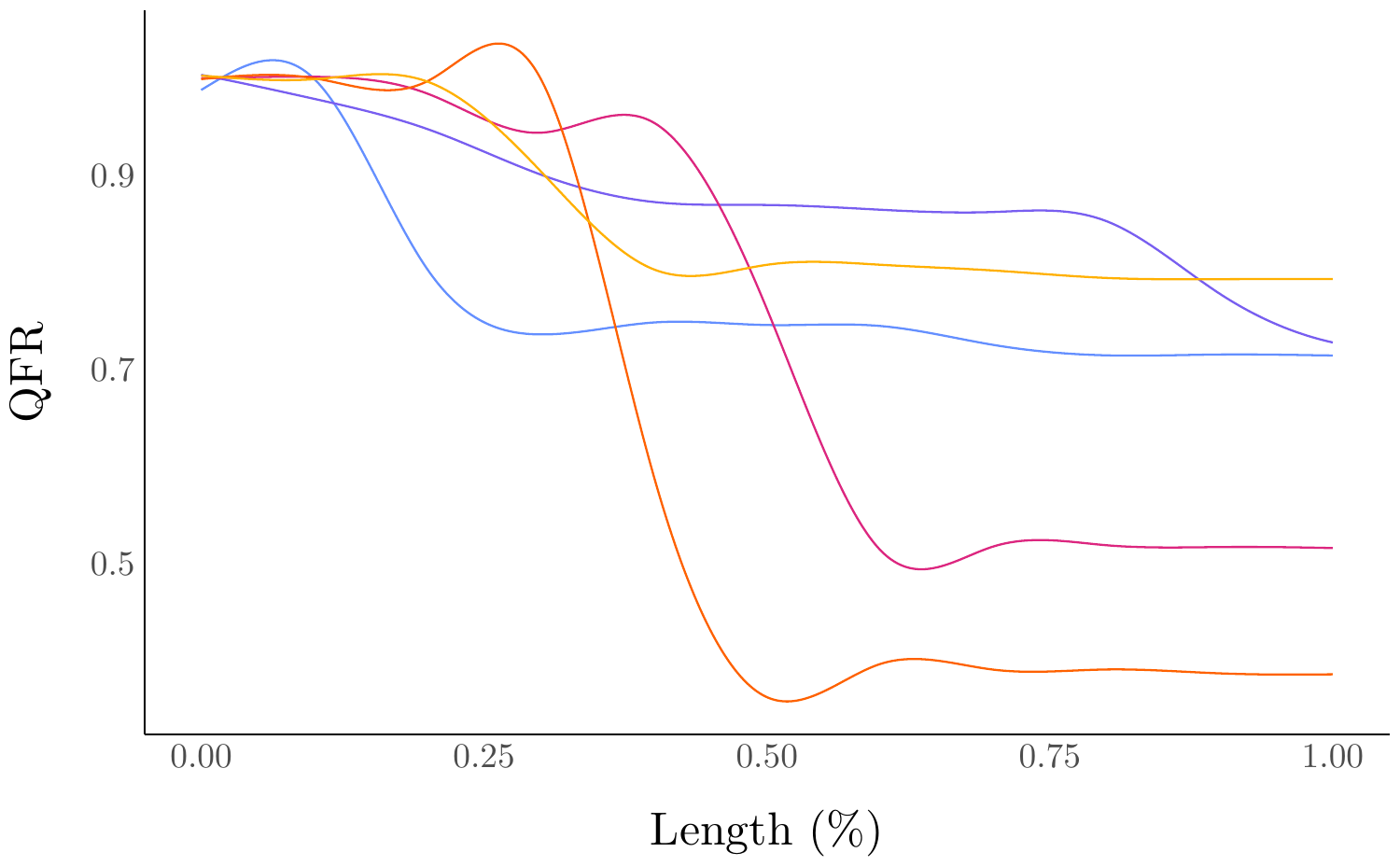}
	\end{subfigure}%
	\caption{A sample of the raw (top panel) and smoothed (bottom panel) diameter and QFR curves.}
	\label{fig:FigAngioData}
\end{figure}

As an illustration, the original trajectories of the diameter and QFR from five vessels are provided in the top panel of Figure~\ref{fig:FigAngioData}, where the x-axis is the length of the vessel. There is a general decreasing trend for both diameter and QFR curves. Specifically, most QFR curves display a rapid decline followed by a relatively steady plateau, whereas the diameter curves exhibit more significant fluctuations along the vessel. We aim to investigate the dynamic changes within the diameter and QFR by recovering their first-order derivatives, and obtaining the functional principal components and scores of the derivatives. Our findings can be helpful for classifying CAD patterns and planning PCI procedures.

The diameter and QFR curves are treated as multivariate functional data. The number of measurement points for each curve varies, ranging from 71 to 319. The data are thus irregular, containing 36,190 unique measurement points for diameter and QFR. This significantly increases the computational cost when estimating the covariance surface. On the other hand, the marginal fluctuations observed particularly in the diameter curves may be attributed to measurement errors. They may adversely affect the recovery of the derivatives. Therefore, we smooth the original data at 1001 equally spaced points over the interval $[0,1]$, using P-splines with the smoothing parameter equal to 0.1. A sample of the smoothed diameter and QFR curves are shown in the bottom panel of Figure~\ref{fig:FigAngioData}. Additionally, the data are mean-centred before running the DMFPCA and DMKL methods.

\begin{figure}[H]
	\centering
	\includegraphics[width=\linewidth]{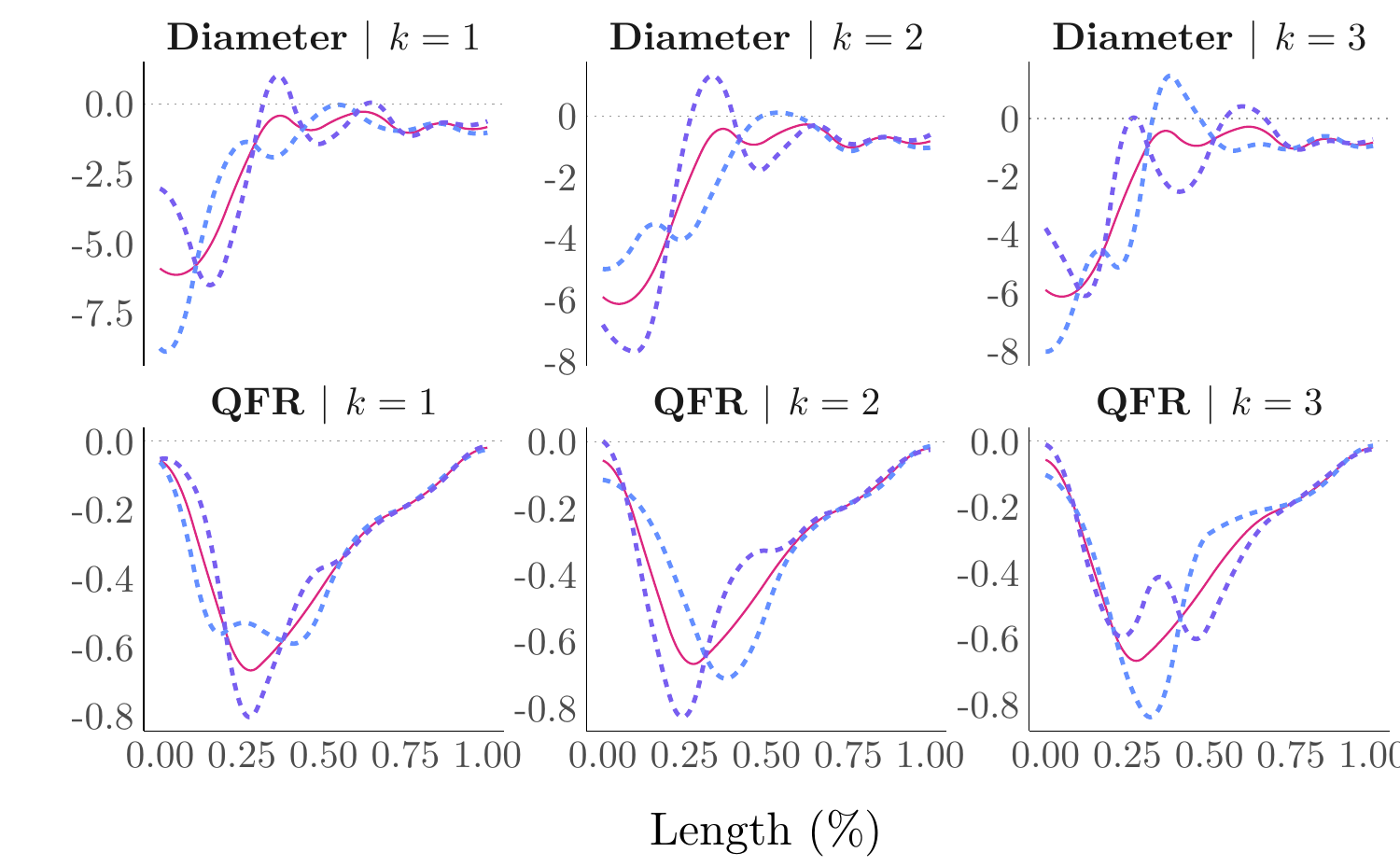}
	\caption{The estimated mean derivative functions (red), plus (blue) and minus (purple) the estimated eigenfunctions by using DMFPCA method.}
	\label{fig:FigDMFPCA}
\end{figure}

\begin{figure}[H]
	\centering
	\includegraphics[width=\linewidth]{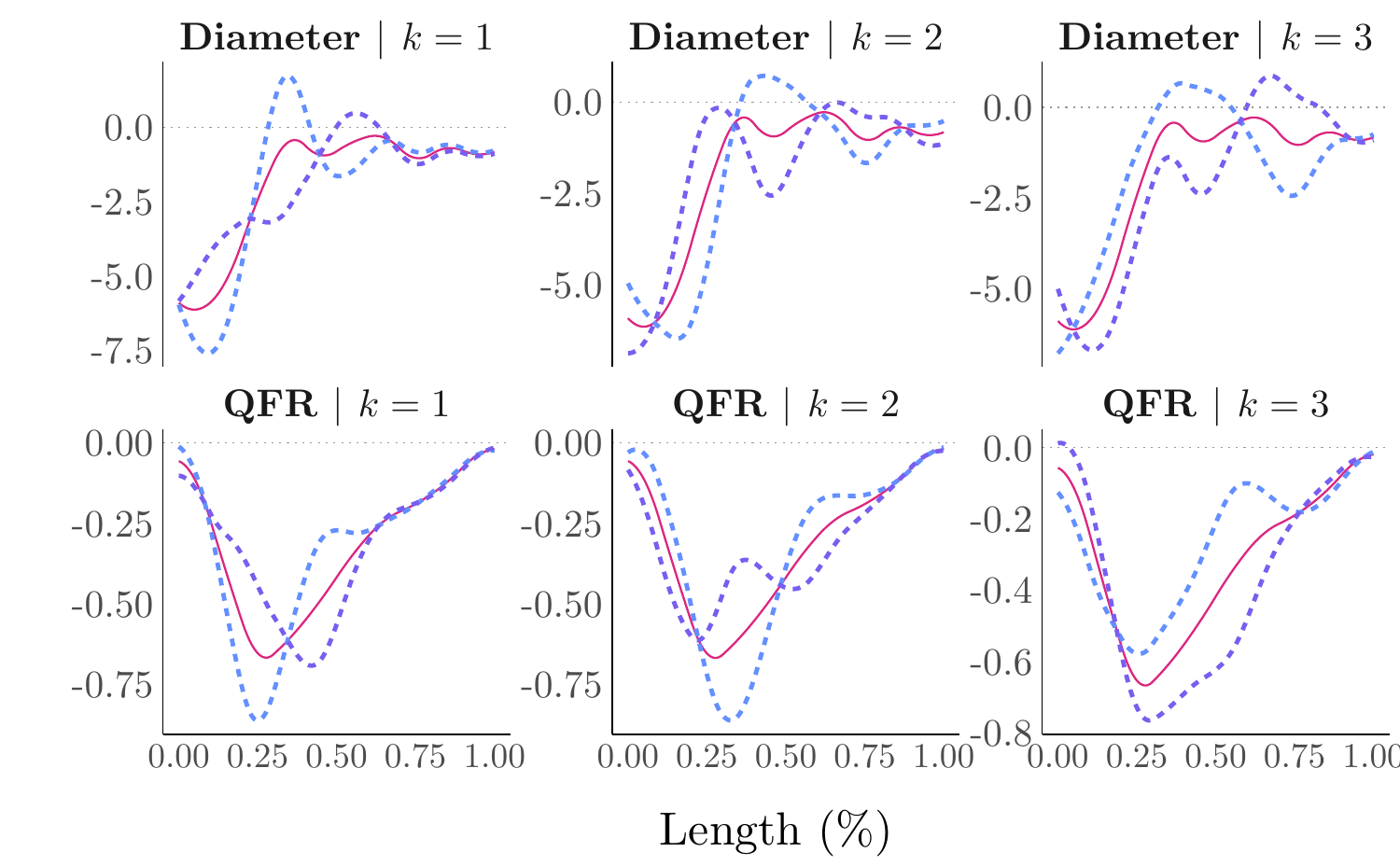}
	\caption{The estimated mean derivative functions (red), plus (blue) and minus (purple) the estimated eigenfunctions by using DMKL method.}
	\label{fig:FigDMKL}
\end{figure}

The mean functions and the corresponding mean derivatives of the diameter and QFR curves are estimated using P-splines. The mean derivatives (red lines) in Figure~\ref{fig:FigDMFPCA} and \ref{fig:FigDMKL} indicate there is an overall decreasing trend for both diameter and QFR throughout the length of the vessel. More specifically, the diameter appears to narrow with a decreasing rate at the initial half of the vessel, and then continues to decrease marginally at a consistent rate towards the end of the vessel. The QFR experiences a rapid decline at the initial segment (around $25\%$) of the vessel and then decreases more gradually towards the end of the vessel. In fact, since the data are collected from patients with coronary artery disease, this condition is likely to cause arterial narrowing and a subsequent decrease in QFR. 

When using DMFPCA and DMKL on the mean-centered data, we set $K=5$ as the truncation number. Figure~\ref{fig:FigDMFPCA} only shows the first three DMFPCs which account for $77\%$ of the total variance. It is observed that for each $k=1,2,3$, the mean derivative of QFR minus the DMFPC (purple line) experiences its minimum around $25\%$ of the vessel, where the mean derivative of diameter minus the DMFPC reaches zero. This indicates that the QFR experiences its fastest drop when the diameter reaches its narrowest point. More specifically, for $k=1$, subtracting the first DMFPC from the mean derivative of QFR results in a single main trough, while adding the first DMFPC gives two troughs that lie within an overall plateau-like region. This suggests that a vessel with a lower first score is more likely to have a focal drop. The second DMFPC affects the point of the narrowest diameter and the fastest QFR drop.  For a vessel with a lower second score, it is more likely to experience this point earlier. The third DMFPC seems to capture a similar variation as the first DMFPC, whereas subtracting the third DMFPC to the mean derivative of QFR gives two sharper troughs. A vessel with a lower third score may experience two drops separated by some distance, which is indicative of serial lesions.

The first three eigenfunctions estimated from DMKL are shown in Figure~\ref{fig:FigDMKL}. The first eigenfunction influences the point of the narrowest diameter and the fastest QFR drop. This implies that a vessel with a higher first score appears to experience this point earlier. For $k=2$, a higher score is more likely to define a focal lesion, whereas a lower score tends to indicate serial lesions. The third eigenfunction affects the location of the narrowest diameter and the scale of the decreasing rate of QFR. A vessel with a higher third score is expected to have a slower rate of QFR decline across a broader diameter.

\begin{figure}[H]
	\centering
	\includegraphics[width=\linewidth]{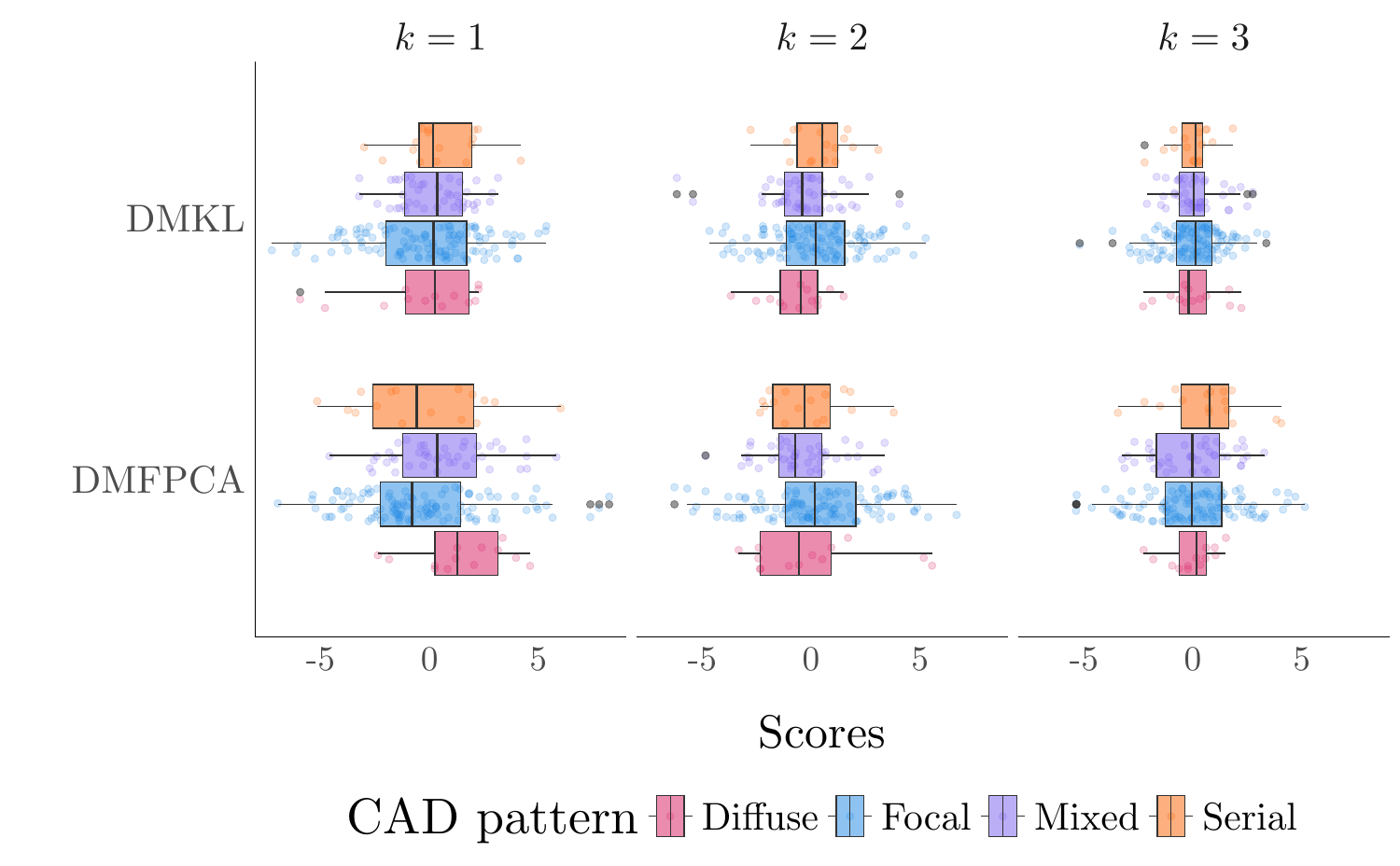}
	\caption{The boxplots of scores corresponding to the first three eigenfunctions from DMKL and DMFPCA.}
	\label{fig:FigAppScores}
\end{figure}

Figure~\ref{fig:FigAppScores} presents boxplots of the scores corresponding to the first three eigenfunctions from DMKL and DMFPCA, grouped by the four CAD patterns. More specifically, we select two vessels with $\text{ID} = 15$ and $\text{ID} = 48$. The estimated scores (first three) for these two vessels are listed in Table~\ref{tab:TabScores}. From DMFPCA, vessel $\text{ID}=15$ has lower score on $k=1$ than $\text{ID}=48$. This indicates vessel $\text{ID}=15$ is more likely to have a focal lesion. Indeed, the CAD pattern of vessel $\text{ID}=15$, as assessed by a panel of cardiologists, is focal. DMKL is less informative, providing the indication that a higher score on $k=2$ for vessel $\text{ID}=15$ corresponds to a focal lesion.

\begin{table}
	\centering
	\caption{The first three scores for vessels with $\text{ID}=15$ and $\text{ID}=48$}
	\begin{tabular}{lllccc}
		\toprule
		ID & Method & CAD Pattern & $k=1$ & $k=2$ & $k=3$ \\
		\midrule
		15 & DMFPCA & Focal & -2.41 & 4.32 & 0.48 \\
		48 & DMFPCA & Serial & 2.17 & 0.26 & 4.07 \\
		15 & DMKL & Focal & -3.16 & 2.29 & 0.88 \\
		48 & DMKL & Serial & -0.34 & 1.12 & 0.03\\
		\bottomrule
	\end{tabular}
	\label{tab:TabScores}
\end{table}

\begin{figure}
	\centering
	\begin{subfigure}{0.49\textwidth}
		\includegraphics[width=\linewidth]{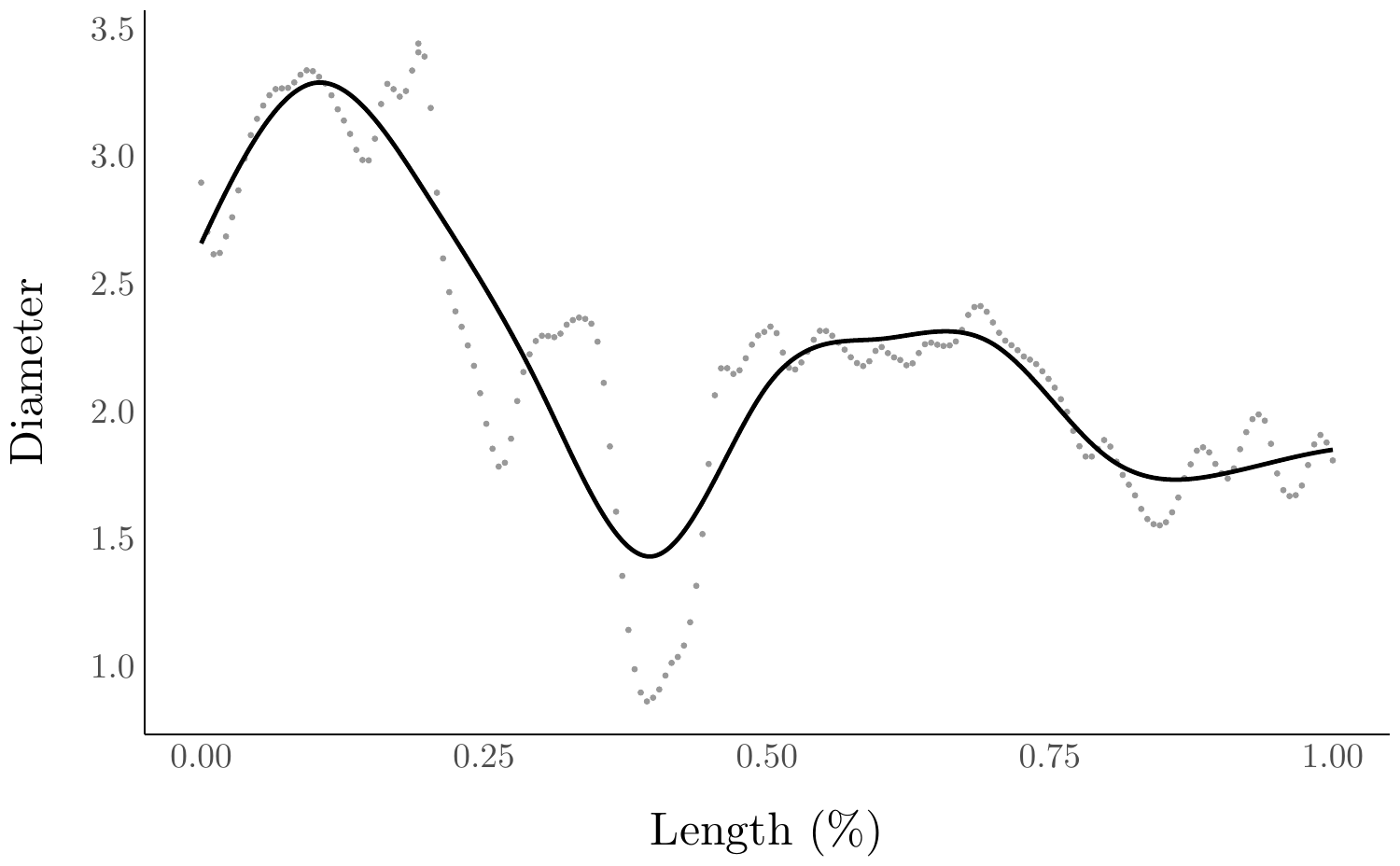}
	\end{subfigure}
	\hfill
	\begin{subfigure}{0.49\textwidth}
		\includegraphics[width=\linewidth]{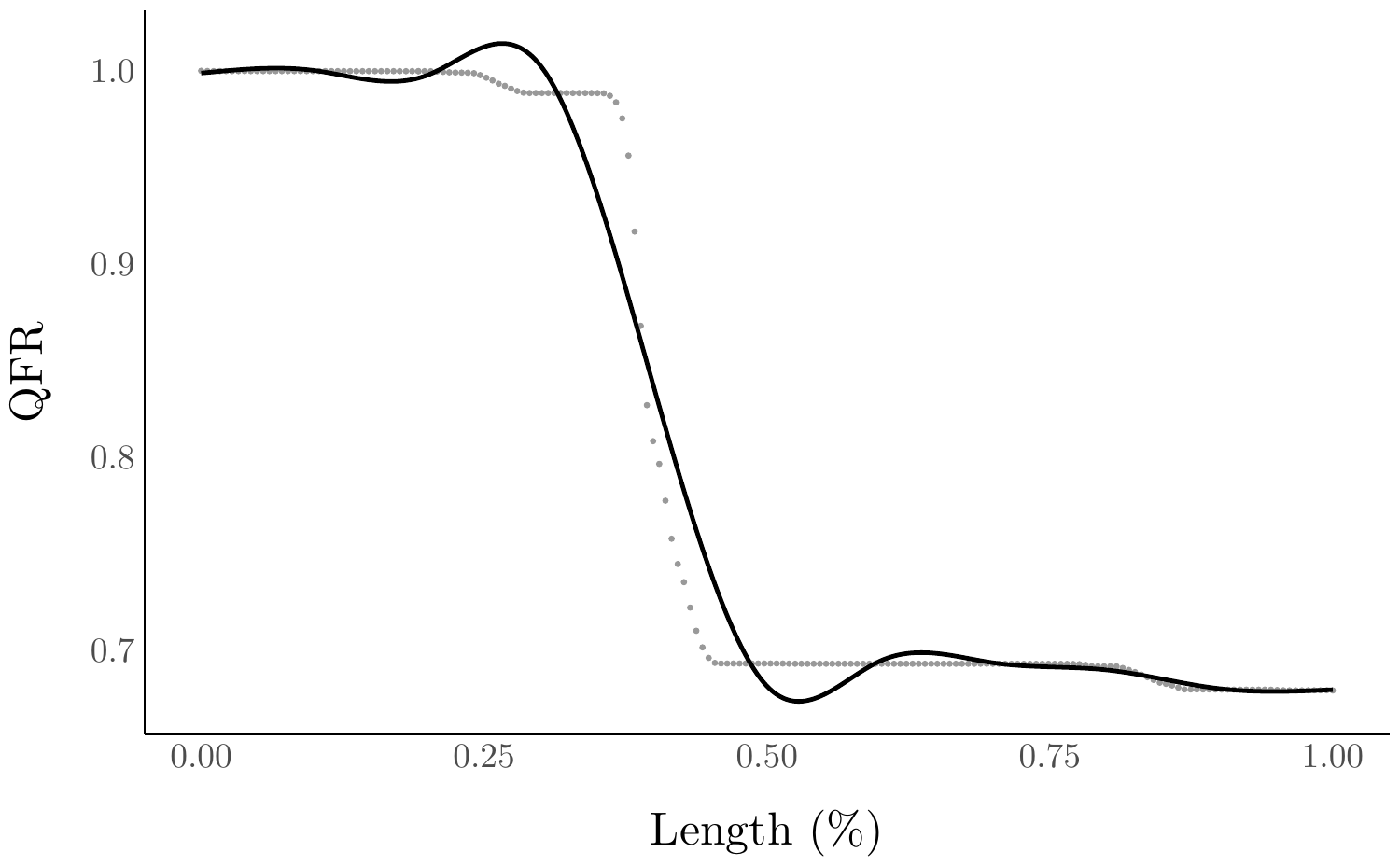}
	\end{subfigure}
	\\
	\begin{subfigure}{0.49\textwidth}
		\includegraphics[width=\linewidth]{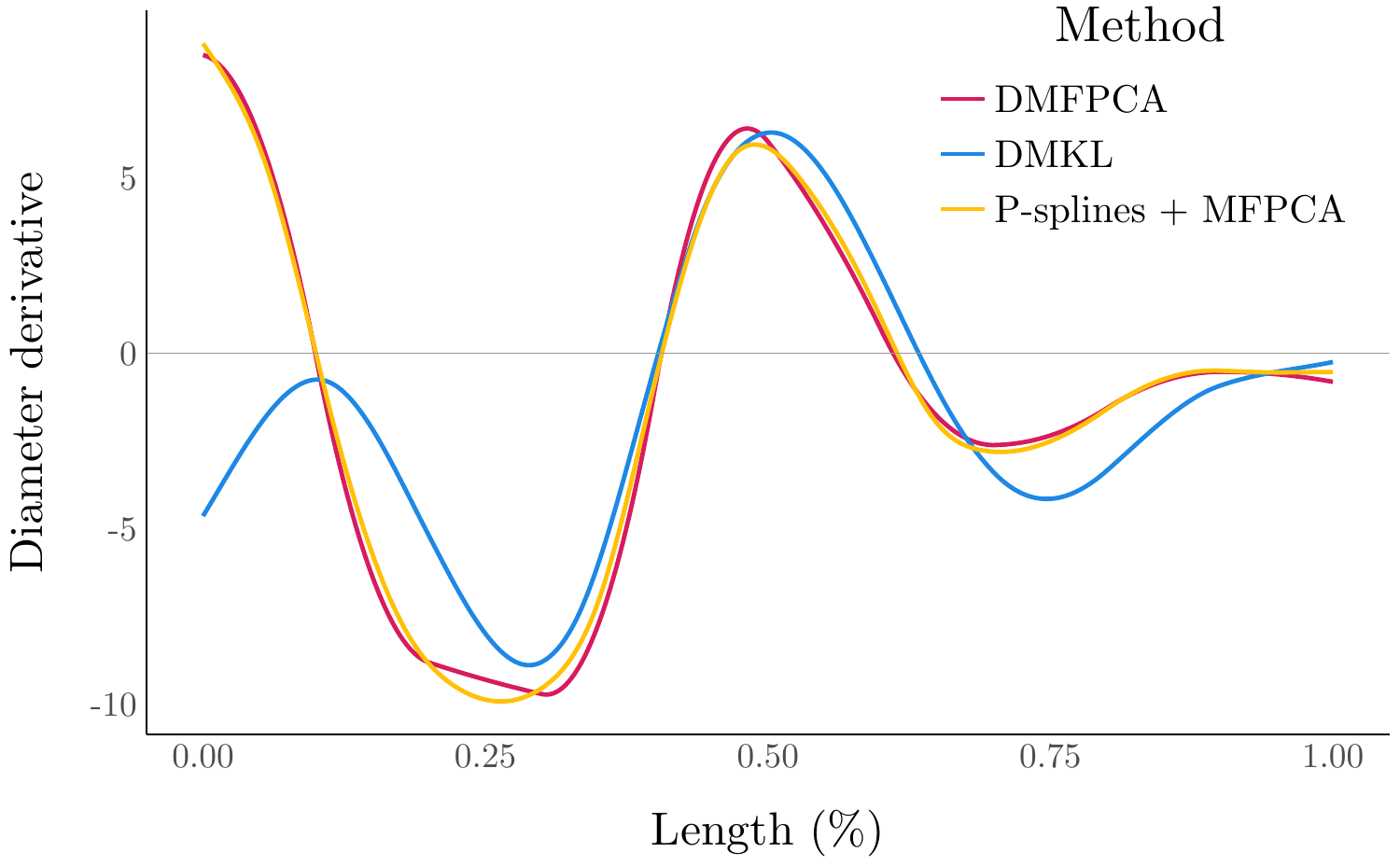}
	\end{subfigure}
	\hfill
	\begin{subfigure}{0.49\textwidth}
		\includegraphics[width=\linewidth]{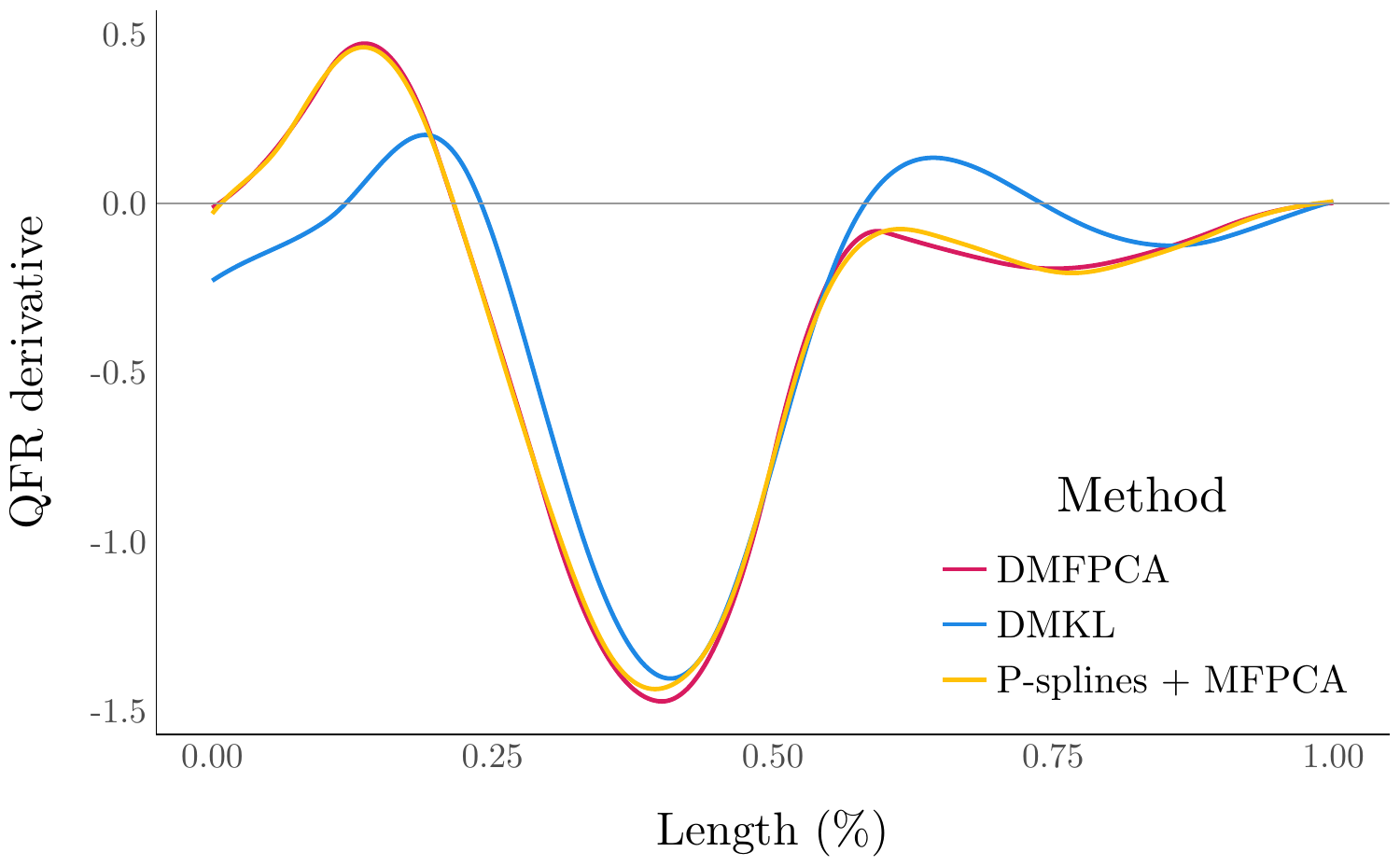}
	\end{subfigure}
	\begin{subfigure}{\textwidth}
		\caption{ID $15$}
	\end{subfigure}
	\\
	\begin{subfigure}{0.49\textwidth}
		\includegraphics[width=\linewidth]{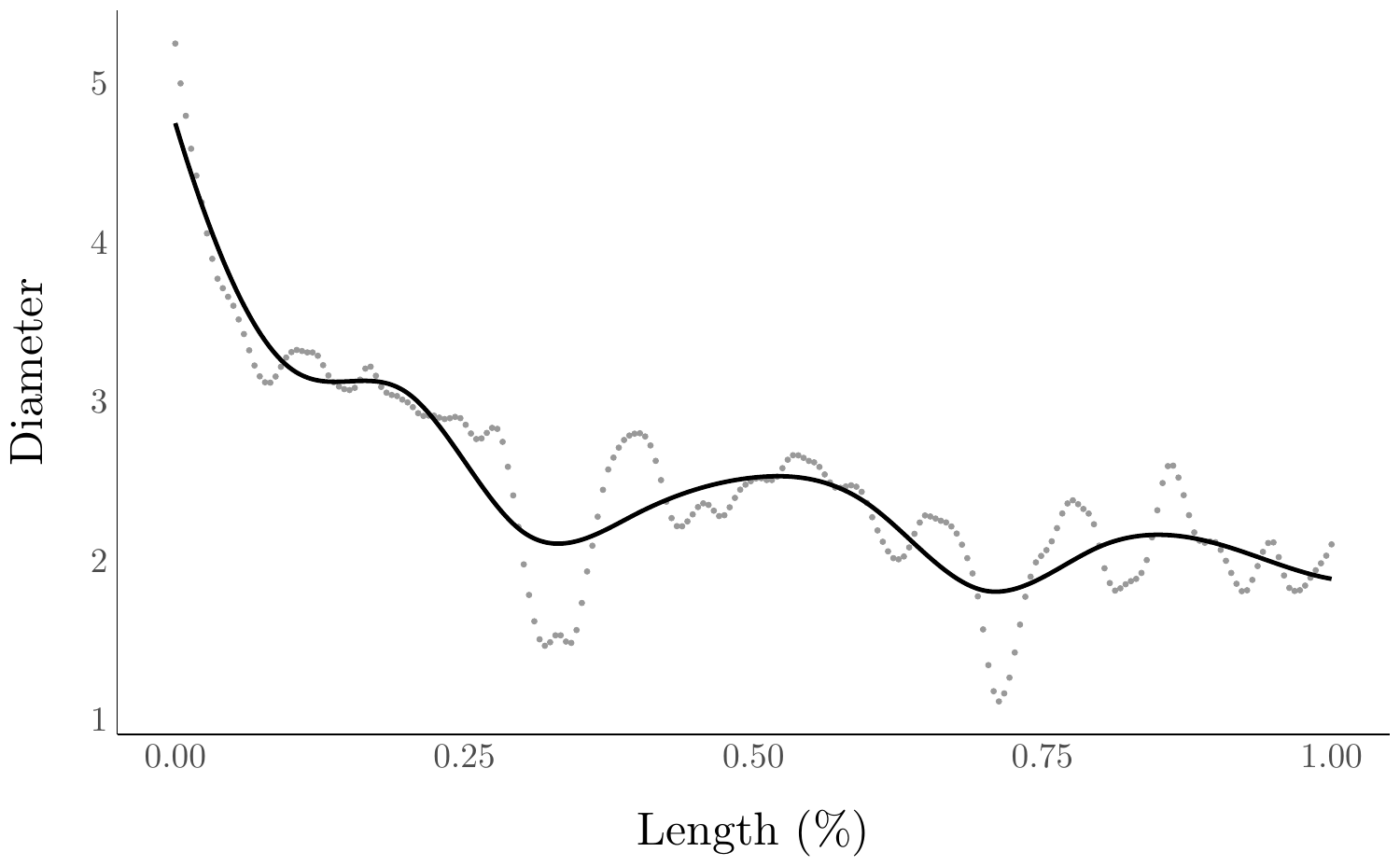}
	\end{subfigure}
	\hfill
	\begin{subfigure}{0.49\textwidth}
		\includegraphics[width=\linewidth]{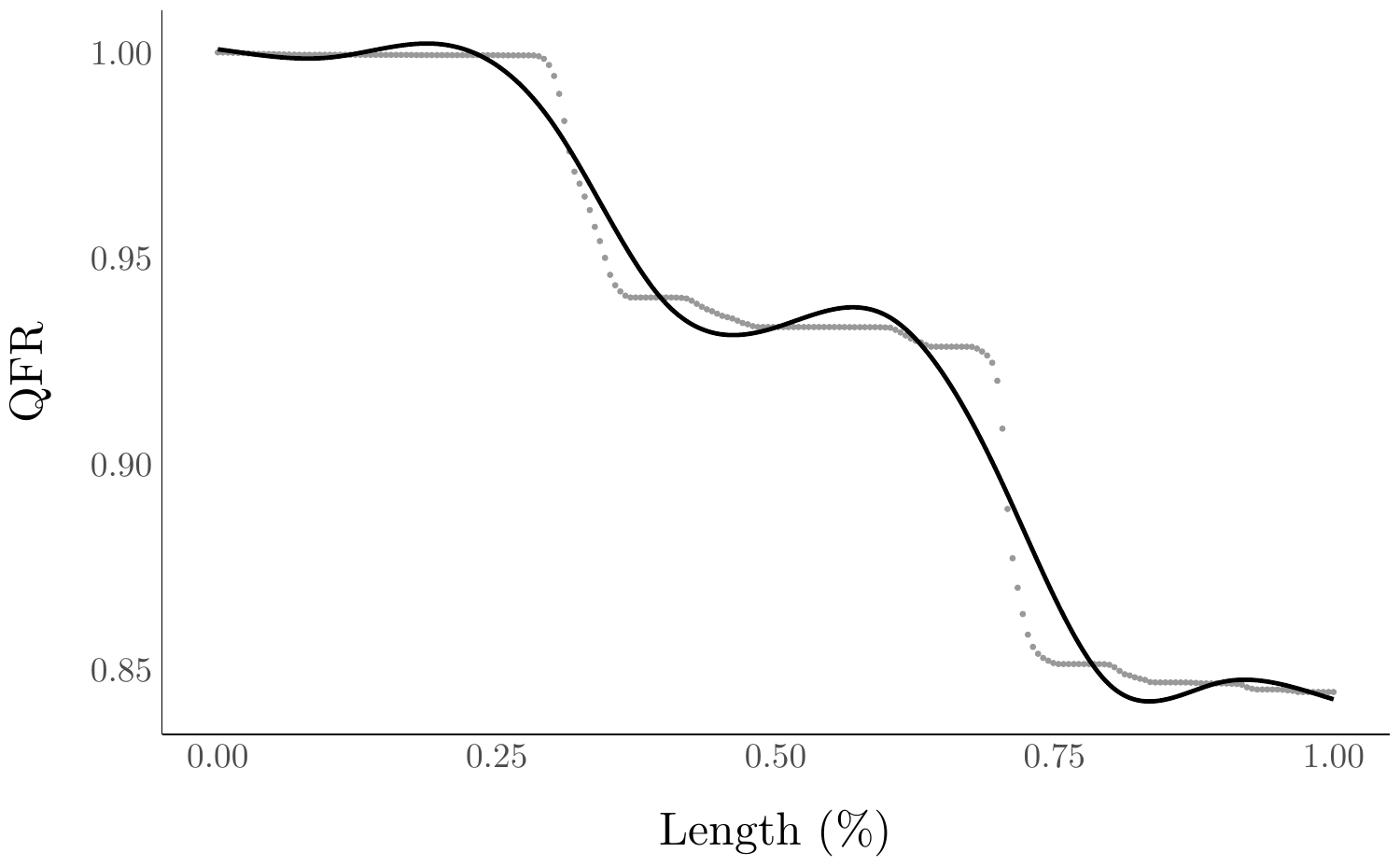}
	\end{subfigure}
	\\
	\begin{subfigure}{0.49\textwidth}
		\includegraphics[width=\linewidth]{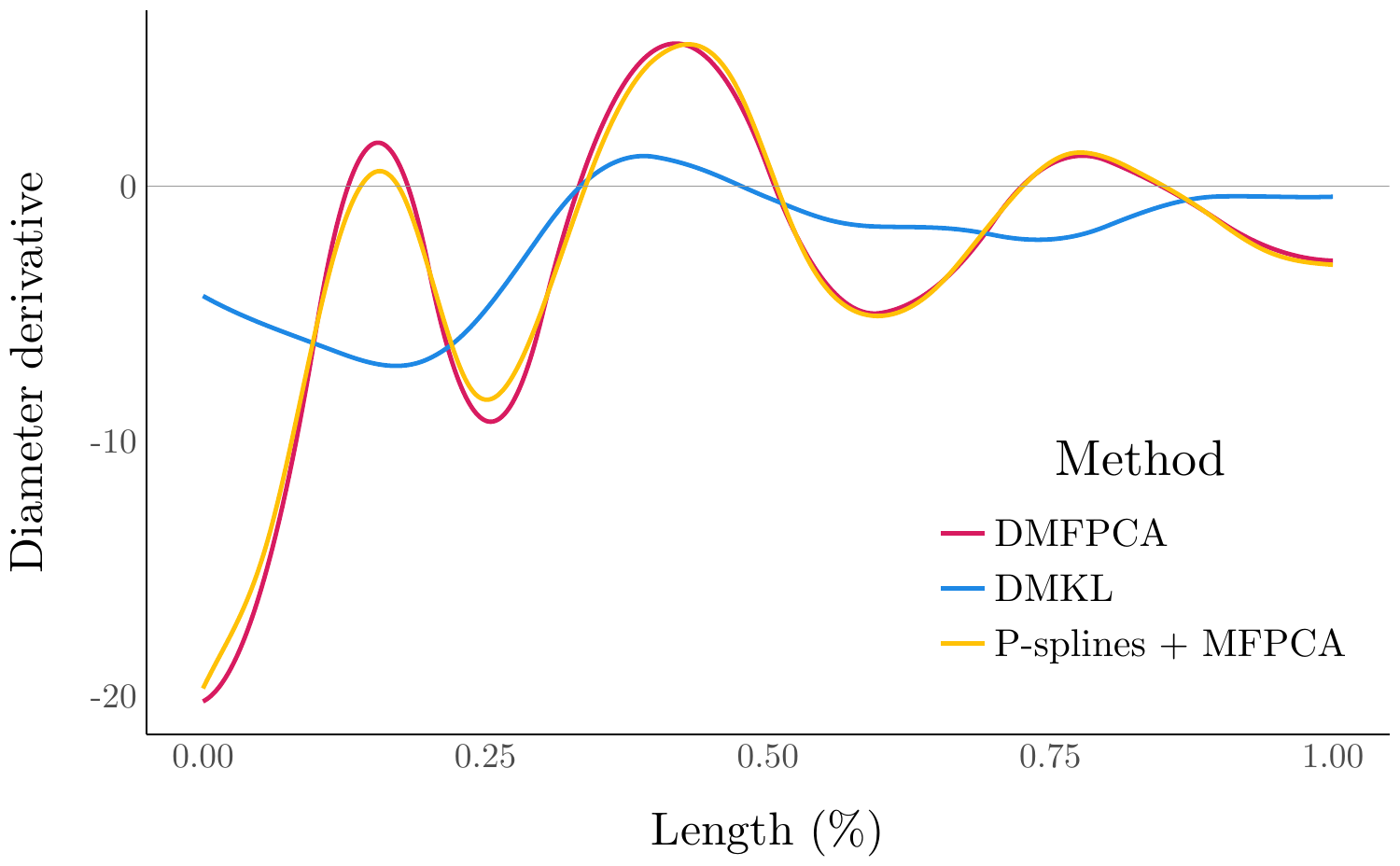}
	\end{subfigure}
	\hfill
	\begin{subfigure}{0.49\textwidth}
		\includegraphics[width=\linewidth]{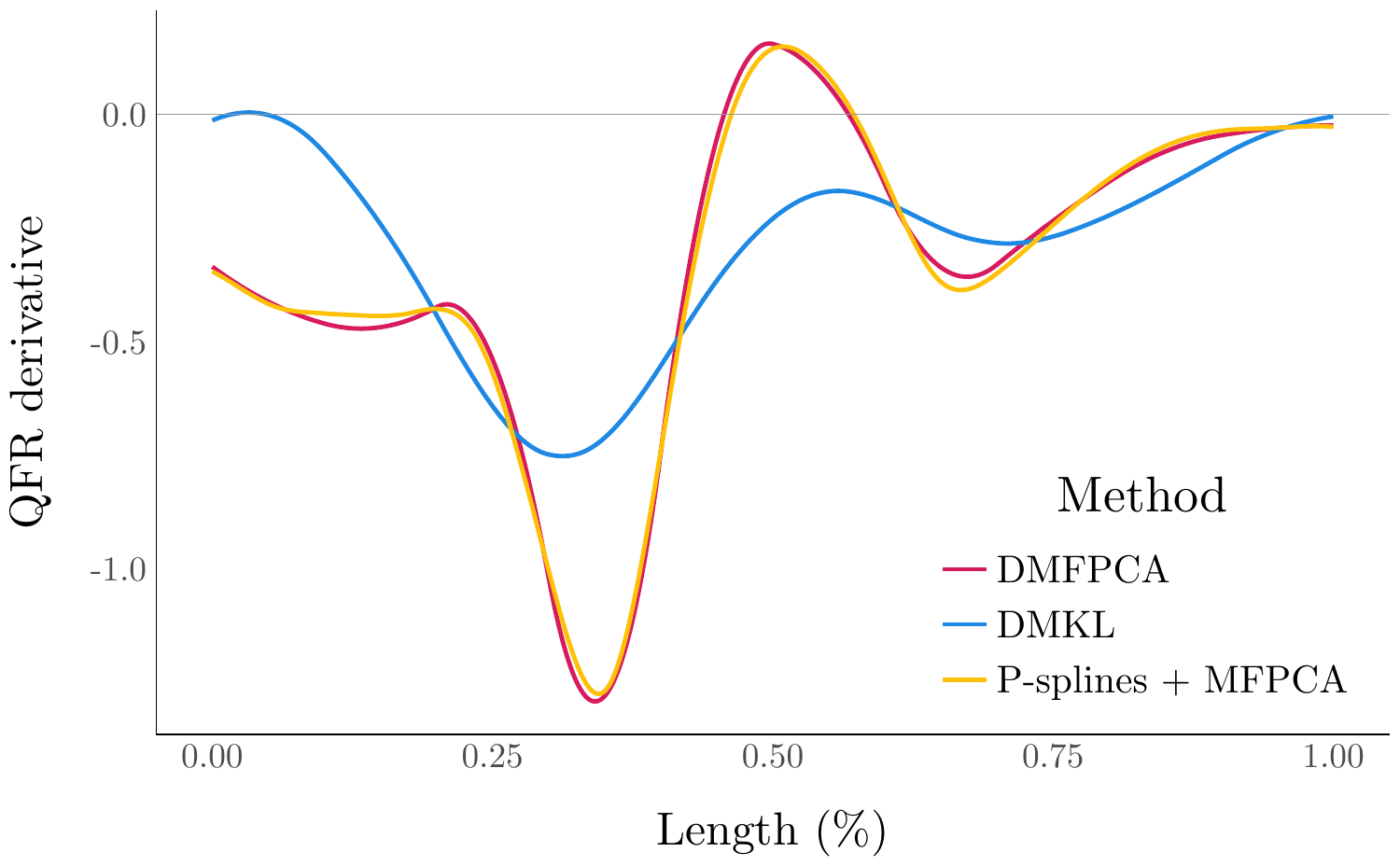}
	\end{subfigure}
	\begin{subfigure}{\textwidth}
		\caption{ID $48$}
	\end{subfigure}
	\caption{Examples of $\text{ID}=15$ and $\text{ID}=48$. The raw data (grey dots) and the smooth data (black lines) are presented in the top panel for each ID. The derivatives refitted by DMFPCA (red lines), DMKL (blue lines) and P-splines + MFPCA (orange lines) are provided in the bottom panel.}
	\label{fig:FigAngioExample}
\end{figure}

Figure~\ref{fig:FigAngioExample} shows the refitted derivatives for the two vessels. For $\text{ID} = 15$, DMFPCA and DMKL provide similar refitted derivatives, except at the very beginning of the vessel. The refitted derivatives of diameter and QFR suggest that the diameter is expected to reach a local minimum at around 40\% of the vessel, near the point where QFR undergoes its steepest drop. Since there is only one significant trough in the refitted derivatives of QFR, the vessel tends to have a focal lesion, which aligns with the cardiologists' assessment. For $\text{ID} = 48$, the refitted derivatives from DMFPCA indicates the diameter experiences its local minima at around 30\% and 75\% along the vessel, which are roughly the points where QFR has the two fastest drops. The two main troughs in the derivatives of QFR indicate the vessel $\text{ID} = 15$ is likely to have serial lesions, which is consistent with the cardiologists' assessment. When using DMKL, the refitted derivatives seem to less accurately capture changes in diameter and QFR near the end of the vessel. For comparison, the method introduced in Algorithm~\ref{alg:derivatives_obs} is also employed, shown as orange lines. It turns out that the refitted derivatives from P-splines + MFPCA  are very close to those from DMFPCA. 

For the angiogram data, DMFPCA appears to provide more informative results than DMKL. The DMFPCs and refitted derivatives effectively reflect the dynamic changes in the diameter and QFR, which would help in detecting the narrowest location with steepest QFR drop. A stent should be placed to such location to reopen a blocked artery and increase the blood flow. More importantly, the DMFPC-scores, which summarise the dynamic changes in the diameter and QFR, would be of interest for further use in classification of CAD patterns.


%% file: main/Discussion.tex
\section{Discussion}
\label{sec:discussion}

In this paper, we propose DMFPCA and DMKL to estimate the eigencomponents and scores of the derivatives for multivariate functional data. The estimated eigencomponents give insights into the modes of variation of the derivatives, and the estimated scores can be used as predictors contributing to further regression analyses. With these two components, the unobserved derivatives can be reconstructed. In general, our methods extend the current derivative estimation approaches which are applied to the univariate functional data to multivariate cases. 

In terms of the estimation procedure, DMFPCA is more straightforward compared to DMKL. The DMKL requires an initial estimation of the derivatives for multivariate functional data. To obtain the initial estimation, MFPCA is firstly applied to the original multivariate functional data, which gives MFPC-scores and MFPCs. By using the derivatives of MFPCs, an initial version of the derivatives for multivariate functional data is estimated. Afterward, MFPCA is applied again, but this time to the initial estimation of the derivatives, which gives the final multivariate eigencomponents and scores. In contrast, when using DMFPCA, no initial estimation of the derivatives for multivariate functional data is needed. The DMFPCA starts with the computation of the derivative of the covariance for each univariate feature. With a spectral decomposition, the univariate eigencomponents are calculated. Afterward, the final multivariate eigencomponents are obtained by linking to their univariate counterparts, based on the relationship introduced in \citet[Prop.~5]{happ2018multivariate}. 

Note that when using DMKL, the derivatives of MFPCs are not orthogonal (see Remark~\ref{rem:dmkl}) and therefore, not the best bases for reconstructing the initial estimation of the derivatives for multivariate functional data. This leads to less accurate estimates of the final multivariate eigencomponents and scores, as indicated from the simulation study. On the other hand, DMFPCs provide the orthogonal eigenfunctions, which depend on the decomposition of the derivative of covariance for each univariate feature. For future work, we could follow \cite{li2020fast} to estimate the derivatives of the covariance functions for multivariate functional data. This would be more complex than univariate cases but allow us to directly target the derivatives of multivariate functional data.

Moreover, a direct approach is introduced in Algorithm~\ref{alg:derivatives_obs}. Analogous to DMKL, it requires an initial estimation of the derivatives, followed by the application of MFPCA to the initial estimation. However, unlike DMKL, the method in Algorithm~\ref{alg:derivatives_obs} estimates the derivative of each curve from different univariate features separately. The choice of the derivative estimation for a single curve is left to the reader (we use P-splines in this paper).

The methods are tested in the simulation study. The multivariate functional data with four features which mimic the biomedical applications are simulated. It turns out that all three methods can provide reliable refitted derivatives, even when the data are highly sparse. However, DMKL may fail to provide accurate estimates of the final eigencomponents. With the application of coronary angiogram data, DMFPCA offers more insights into dynamic changes in the diameter and QFR. The DMFPCA-scores can be further used for CAD patterns classification. Overall, DMFPCA should be the preferred choice over the other two methods, particularly for practical applications.
